\documentclass{emulateapj}
\usepackage[T1]{fontenc}
\usepackage{tgtermes}
\usepackage{mathptmx}  
\usepackage{amsthm,amsmath,amssymb}
\usepackage{mathrsfs}
\usepackage[usenames, dvipsnames]{color}
\usepackage{hyperref}

\newcommand{\Brian}[1]{{\color{OrangeRed} #1}}

\newcommand{\Del}[1]{{\color{ForestGreen} #1}}
\newcommand{\Sam}[1]{{\color{BurntOrange} #1}}

\begin{document}

 \title{Mapping Spatial Variations of HI Turbulent Properties in the Small and Large Magellanic Cloud}
  \shorttitle{Spatial variations of HI turbulent properties across the SMC and LMC}

\author{Samuel Szotkowski$^1$, Delano Yoder$^1$, Sne\v{z}ana Stanimirovi\'{c}$^1$, Brian Babler$^1$, N. M.  McClure-Griffiths$^2$, Helga D\'{e}nes$^3$, Alberto Bolatto$^4$,  Katherine Jameson$^2$, Lister Staveley-Smith$^5$}
\affil{$^1$ Astronomy Department, University of Wisconsin - Madison, 475 North Charter Street, Madison, WI, 53706-1582USA, $^2$ Research School of Astronomy \& Astrophysics, Australian National University, Weston Creek, ACT, Australia, 
$^3$ ASTRON, The Netherlands Institute for Radio Astronomy, Dwingeloo, Netherlands,
$^4$ Department of Astronomy, University of Maryland, College Park, MD, 20742, USA, $^4$ ASTRON, The Netherlands Institute for Radio Astronomy, Dwingeloo, Netherlands, $^5$ ICRAR, University of Western Australia, Crawley, Perth, WA, Australia}

\begin{abstract}
 We developed methods for mapping spatial variations of the spatial power spectrum (SPS) and structure function (SF) slopes, with a goal of connecting neutral hydrogen (HI) statistical properties with the turbulent drivers. The new methods were applied on the HI observations of the Small and Large Magellanic Clouds (SMC and LMC). 
 In the case of the SMC, we find highly uniform turbulent properties of HI, with no evidence for local enhancements of turbulence due to stellar feedback.
Such properties could be caused by a significant turbulent driving on large-scales.
Alternatively, a significant line-of-sight depth of the SMC could be masking out localized regions with a steeper SPS slope caused by stellar feedback. In contrast to the SMC, the LMC HI shows a large diversity in terms of its turbulent properties. Across most of the LMC, the small-scale SPS slope is steeper than the large-scale slope due to the presence of the HI disk.
On small spatial scales, we find several areas of localized steepening of the SPS slope around major HII regions, with the 30 Doradus region being the most prominent. This is in agreement with predictions from numerical simulations which suggest steepening of the SPS slope 
due to stellar feedback eroding and destroying interstellar clouds. We also find localized steepening of the large-scale SPS slope in the outskirts of the LMC. This is likely caused by the flaring of the HI disk, or alternatively  ram-pressure stripping of the LMC disk due to the interactions with the surrounding halo gas.
\end{abstract}

\section{Introduction}
	Turbulence is an important structuring agent in the interstellar medium (ISM), thought to be responsible in part for formation of molecular clouds as well as complex structure found in spectral-line images \citep{Elmegreen04}.  Many observational studies have revealed a fractal (self-similar, scale-free) nature in distributions of interstellar gas and dust, which when compared with theory and terrestrial turbulence experiments indicate that turbulence is an important process in the ISM.  This turbulence is often characterized by estimating spatial power spectrum (SPS) of intensity fluctuations, relating those to the power spectra of the underlying velocity or density fluctuations, and illuminating the spatial scales at which energy is injected or dissipated, as well as the mechanisms by which that energy cascades from large to small scales or vice versa.   Many physical processes including gravitational instabilities \citep{Wada02,Bournaud10,Krumholz16} and stellar feedback \citep{Kim01,deAvillez05,Joung06} have been considered as possible drivers of interstellar turbulence, however, their relative importance remains an open question. 
    
    Stellar feedback, in particular, has been considered as a key agent for establishing the hierarchy of structures in the ISM. For example, \cite{Grisdale17} run numerical simulations and showed that feedback both removes gas and destroys interstellar and giant molecular clouds, effectively shifting the power from small to large scales. In particular, they simulated galaxies like the Small and Large Magellanic Clouds (SMC and LMC), with and without the stellar feedback included, and showed the change of the SPS slope on both small and large scales when feedback is included. In a followup study,
    \cite{Grisdale18} investigated the places where feedback has the first impact -- Giant Molecular Clouds (GMCs) -- and concluded that including effective feedback in numerical simulations is necessary to reproduce observed properties of GMCs.
    
    In this study we continue our investigation of spatial variations of turbulent properties as a way of quantifying the importance of stellar feedback, gravity and ram pressure as structuring agents of the neutral hydrogen (HI) distribution in galaxies. We focus here on the SMC and LMC as they are especially handy targets due to being closeby (50 kpc for the LMC and 60 kpc for the SMC Westerlund 1991), with a wealth of high-resolution observations and ability to probe large spatial dynamic ranges. The SMC and the LMC are part of a three galaxy interacting system with the Milky Way (MW) being the most massive player.  Both the LMC and the SMC have a smaller mass and a lower metallicity than the MW, 1/2 Solar for the LMC and 1/5 Solar for the SMC, 
    \citep{Russell92,Dufour84}.

    In \cite{Nestingen-Palm17} we compared HI turbulent properties between the main star-forming body and outskirts of the SMC. We did not find any significant difference either when probing mainly the density or velocity fluctuations, suggesting that large-scale turbulent driving may be more prominent than stellar feedback in the SMC. In the present study we develop a new technique for mapping out the slope of the HI turbulent spectrum across the SMC and LMC. Our goal is to connect turbulent properties with the underlying physical processes.

    Turbulent properties of HI and dust in the SMC and LMC have been investigated in numerous prior studies. The HI SPS of the SMC has been known not to show any breaks or turnovers \citep{Stanimirovic99,Stanimirovic01}, suggesting significant energy injection on spatial scales larger than the size of the SMC. The LMC HI SPS, however, has shown a break at 100-200 pc, which was interpreted as being due to disk geometry \citep{Elmegreen00a,Padoan01}. Similar findings have been found for the dust column density, no breaks in the case of the SMC \citep{Stanimirovic00} and a clear break in the case of the LMC \citep{Block10}.  In the case of the SMC, the dust column density SPS slope agreed with the SPS slope of the HI column density, supporting the idea that gas and large dust grains are well mixed in the ISM. 
    In the case of Spitzer observations of the LMC, the SPS slopes were found to be slightly steeper for longer wavelengths, suggesting that cooler dust emission is smoother than the hot dust emission (Block et al. 2010). 
    
    Several additional studies have found breaks in the SPS \citep{Elmegreen03,Dutta09a,Dutta09b}.
\cite{Combes12} found a SPS slope of $-1.0$ to $-1.5$ on spatial scales $>500$ pc in M33, while a slope of $-3.0$ to $-4.0$ on smaller spatial scales, with a break at 100-200 pc observed using several tracers.
    Also, the HI SPS slope has been shown to depend on the thickness of velocity channels, with narrow channels probing largely velocity fluctuations while thick channels probing largely density fluctuations \citep{Lazarian99,Stanimirovic01}. In addition, \cite{Lazarian04} showed theoretically that the SPS slope can be affected by optical depth. While in the SMC the fraction of cold HI is small, $\sim20$\% \citep{Jameson19}, and the optical depth correction is not significant \citep{Nestingen-Palm17}, 
    the LMC has the CNM fraction of $\sim33$\% \citep{Dickey00} and likely a more significant correction for high optical depth. 
    While \cite{Braun12} estimated a correction factor of 1.3 for the LMC based on HI emission spectra alone,
    understanding fully the magnitude of the optical depth correction will require future HI absorption observations 
    for many radio sources behind the LMC.
    
    From early studies it was apparent that the SMC and LMC HI SPS slopes are different. \cite{Muller04} also noticed significantly different SPS slopes for four regions in the Magellanic Bridge between the SMC and LMC. These differences of turbulent properties on kpc-scales in the Bridge were interpreted as likely tracing different origin (age and physical properties) of gaseous arms pulled out of the SMC. Some small spatial variations of HI turbulent properties across the SMC were noticed by \cite{Burkhart10} where skewness and kurtosis were used to estimate the sonic Mach number based on a relationship found in numerical simulations. HI observations of many other nearby galaxies were used to study turbulent properties via fitting of averaged HI velocity profiles 
    \citep{Young96,Warren12,Tamburro09,Stilp13}. It was commonly found that the 
    linewidth of broad
    HI components decrease with galactocentric radius. One interpretation of this trend is the decrease of turbulent energy due to lack of stellar feedback. 
     The Burkhart et al. (2010) method was also applied to a set of  spiral  galaxies  from  The  HI Nearby  Galaxy  Survey (THINGS) in    \cite{Maier16}. Generally, uniform statistical
moments were found across galaxies without an obvious correlation between moments and star-forming regions. However, this study had a resolution of about 700 pc and the statistical moments are likely tracing only large-scale turbulence.  
While \cite{Dutta09b} hinted that dwarf galaxies with higher star formation rate (SFR) per unit area have steeper HI SPS based on a sample of seven galaxies,
\cite{Zhang12} 
found a lack of correlation between the SPS slope and the SFR surface density for
a sub-sample of LITTLE THINGS dwarf irregular galaxies.

This paper is organized in the following way. In Section~\ref{s:data} we summarize new and existing data sets used in this study. Section~\ref{s:methods} explains our rolling SPS and structure function (SF) approach for mapping out spatial variability of turbulent properties across the SMC and LMC. We also summarize here how we produce simple images where we know the SPS slope to test and constrain biases involved with out rolling methods. Several of our tests are documented in the Appendix section. Sections~\ref{s:SMC} and \ref{s:LMC} present our results and compare SPS/SF slopes with the star formation rate and stellar surface densities. 
Finally, we discuss and conclude our results in Section~\ref{s:discussion}.

\section{Data}
\label{s:data}

\subsection{SMC and LMC HI Observations}

\subsubsection{ATCA+Parkes HI observations of the SMC and LMC}

	The neutral hydrogen (HI) data cube of the SMC used in this study is from \cite{Stanimirovic99}. 
    This is a combination of an aperture synthesis mosaic obtained with the Australia Telescope Compact Array (ATCA)
    and single-dish observations obtained with the Parkes 64-m telescope. The two data sets were combined in the image domain, the restoring beam size is $98"$. The cube contains 78 velocity channels, each with a velocity resolution of 1.65 km/s. The HI column density noise level is $4.2 \times 10^{18}$ cm$^{-2}$ per 1.65 km/s velocity channel.
For further details please see Stanimirovic et al. (1999). 
    
 The HI data cube for the LMC is from \cite{Kim03}.
 The Parkes HI observations with the multi-beam receiver \citep{Staveley-Smith03} were combined with the ATCA aperture synthesis mosaic from Kim et al. (1998), resulting in 
the final data set with the angular resolution of $1'$ and the column density sensitivity of $7.2 \times 10^{18}$ per 1.65 km/s velocity channels. The image combination of interferometric and single-dish data was performed in the Fourier domain, using Miriad's task IMMERG. For further details please see \cite{Kim03}.

\subsubsection{ASKAP+Parkes HI observations of the SMC}

   This study also includes recently released SMC HI data from the Australian Square Kilometer Array Pathfinder (ASKAP) as part of Commissioning and Early Science observations. ASKAP is an interferometer with 36 12-m dishes, each with a phased array feed. These observations used only 16 ASKAP antennas \citep{McClure-Griffiths18}. The total integration time was 36 hours in one continuous pointing. ASKAP data were calibrated using the ASKAPSoft pipeline \citep{Cornwell11} and imaged with Miriad. The ASKAP data were combined in the Fourier domain with Parkes data from the HI4PI survey \citep{HI4PI16}. 
   The column density sensitivity of the combined dataset is $5.0 \times 10^{18}$ per 3.9 km/s velocity channels, which is a factor of two lower than that of the ATCA+Parkes data cube.
   With the beam size of $27" \times 35"$, this data set is sensitive to linear scales from 10 pc to 5.3 kpc.

\begin{figure}[h!]
  \centering
  \includegraphics[width=0.5\textwidth]{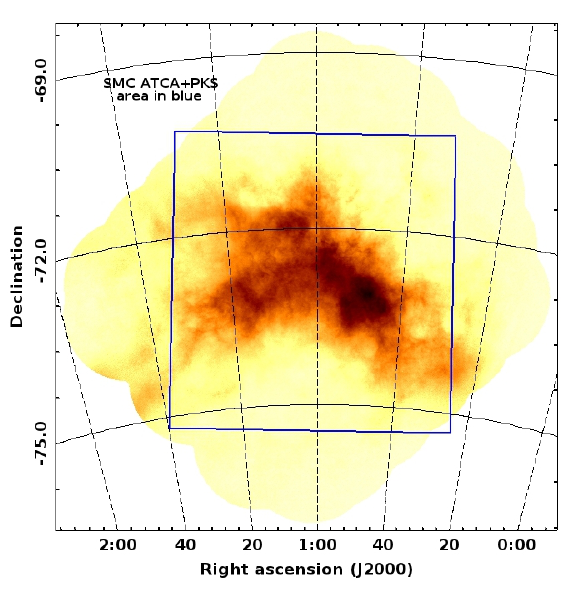}
  \caption{The HI integrated intensity image of the SMC obtained with ASKAP + HI4PI data \citep{McClure-Griffiths18}. To produce this image the HI data cube was  integrated over the velocity range of (46.5 to 238.0 km/s). The angular resolution is $27" \times 35"$. The footprint of the SMC HI ATCA + Parkes data from Stanimirovic et al. (1999) is shown in blue.}
  \label{f:SMC-ASKAP-img}
\end{figure}

\subsection{SFR}

Following \cite{Bolatto11} and \cite{Jameson16}, we combine H$\alpha$ and 24 $\mu$m images to trace recent star formation. The H$\alpha$ image of the SMC is from the Magellanic Cloud Emission Line Survey (MCELS; \cite{Smith99}), and has a resolution of $2.3''$. The H$\alpha$ image of the LMC is from the Southern H$\alpha$
Sky Survey Atlas \citep{Gaustad01}, and has a resolution of $0.8'$. For both the SMC and LMC, the 
24 $\mu$m images are from the Spitzer Survey ``Surveying the Agents of Galaxy Evolution'' (SAGE; \cite{Gordon11}).  To estimate the star formation rate surface density we follow the prescription by \cite{Calzetti07}:
\begin{equation}
SFR(M_\odot\ yr^{-1})=5.3 \times 10^{-42}[L(H{\alpha})+(0.031{\pm}0.006)L(24 {\mu}m)]
\end{equation}
where $L(H{\alpha})$ and $L(24 ~{\mu}m)$ are the luminosities of H$\alpha$ and 24 $\mu$m images in units of erg s$^{-1}$. The SFR image has  the rms noise value
of $4 \times 10^{-4}$ M$_{\odot}$ yr$^{-1}$ kpc$^{-2}$ and $1 \times 10^{-4}$ M$_{\odot}$ yr$^{-1}$ kpc$^{-2}$, for the SMC and LMC respectively.

\section{Methods}
\label{s:methods}

\subsection{Spatial Power Spectrum (SPS)}

We apply a methodology similar to previous studies \citep{Crovisier83,Green93,Stanimirovic99,Elmegreen00}. For a 2D image  $I(x)$, the 2D spatial power spectrum is defined as:
\begin{equation}
P(k) = \int \int \langle I(x) I(x') \rangle e^{i \mathbf{L} \cdot \mathbf{k}} ~d\mathbf{L}
\end{equation}
where $k$ is the spatial frequency, measured in units of wavelength and being proportional to the inverse of the spatial scale, while $\mathbf{L = x-x'}$ is the distance between two points.

We first regrid the HI images to ensure independent pixels, with each pixel having an angular size equal to that of the telescope beam. 
Then we take the Fourier transform and square the modulus of the transform,
$\langle \Re^2+ \Im^2 \rangle$ (where $\Re$ and $\Im$ are the real and imaginary parts of the 2D Fourier transform), which we refer to as the modulus image. A range of annuli is then selected in the modulus image, spaced apart evenly by $0.05 \log(k)$. 
Within each annulus we assume azimuthal symmetry and calculate the median value. As shown in \cite{Pingel13}, median is a better representation of the average power due to the presence of occasional bright pixels in the modulus image 
caused by the Gibbs phenomenon. 
The Gibbs phenomenon (or the edge effect) results from image edges as galaxies are extended and HI column densities never reach zero at the edges of the surveyed regions.  As the Fourier transform of a step function is a sinc function \citep{Bracewell06}, bright pixels appear at the center of $P(k)$ (an example is shown in Figure 2 of Pingel et al. 2018).
The median power is then plotted as a function of $\log$ (spatial scale), and the uncertainties are estimated using the median
absolute  deviation (MAD), e.g. \cite{Pingel18}.

In the SMC, it was found that a single power-law function fits well the HI SPS, $P(k) \propto k^{-\gamma}$ \citep{Stanimirovic99,Stanimirovic01}.
In the LMC however, the SPS was best fit by a double power-law Elmegreen (2001), with the ``break point'' corresponding to a physical scale of $\sim100$ pc.  To estimate the break point and the SPS slopes, in our fitting routine we iterate over all bins in the SPS (excluding 15\% of the bins at the largest and smallest scales), fit a double power law with each bin being selected as the break point.  We then calculate the $\chi^2$ statistics and select as the best fit the break point that minimizes $\chi^2$. We tested our fitting 
procedure on the HI column density subregions that were analyzed by Elmegreen (2001), 
and also on the infrared 160 $\mu$m image from the Spitzer Space Telescope used in \cite{Block10}.  
In both analyses we found breaks, as well as SPS slopes, in agreement with results from those papers. This demonstrates that our fitting routine is reliable.

\subsubsection{The Rolling SPS}

In all previous studies a single or double power-law SPS slope was fitted for the entire HI images of the SMC and LMC,  or several large subregions. Motivated by the improved spatial resolution of ASKAP data, we investigate spatial variations 
of the SPS slope $\gamma$ on much smaller scales than was previously done.  To do this we apply a rolling SPS method. We extract a square sub-image of the HI column density image  (850 x 850 pc for the SMC and 1.5 x 1.5 kpc for the LMC), calculate $\gamma$ at large and small scales along with the corresponding break point. We then shift the sub-image by one beam width (10 pc for the SMC and 14.5 pc for the LMC) and repeat the procedure until the slope and the break point are calculated  across the entire image. During the SPS fitting process, we take the average MAD value of a given bin across all sub-images as the uniform weight given to that bin. We do this because sometimes the largest-scale bin has only few data points, all with similar values, giving that bin extremely large weight. 

In addition, for each kernel where we calculate the SPS, we test whether a single or a broken power law fits the SPS better, in a statistically significant manner, by employing  the $F$-test \citep{Snedecor89}.
We estimate the goodness of fit for both models by the chi-squared statistic:
\begin{equation}
\chi^2 = \sum_{i}^{n}\frac{(Y_{obs, i}-Y_{calc, i})^2}{\sigma_i^2}
\end{equation}
where $n$ is the number of bins in the SPS, $Y_{obs, i}$ is the observed power in a given bin, $Y_{calc, i}$ is the power according to the model, and $\sigma_i^2$ is the variance within the bin.  The $\chi^2$ values for the two models are then used to calculate the $F$-value:
\begin{equation}
F = \frac{(\chi^2_1-\chi^2_2)/(p_2-p_1)}{\chi^2_2/(n-p_2-1)}
\end{equation}
where the subscripts 1 and 2 refer to the single power-law and double power-law models, respectively, and $p$ is the number of parameters in either model.  At a significance level of 5\%, a value $F_{crit}$ can be calculated from a $F(p_2-p_1,n-p_2-1)$ distribution such that the null hypothesis of a single slope fit is rejected if $F>F_{crit}$ (Snedecor \& Cochran 1989, 343-347).


To test the validity and limitations of our rolling SPS method, we applied the same technique to simulated data with known $\gamma$.  We simulated three images:
two with a single input SPS slope of $-2.7$ and $-3.7$, and one with  a double power-law SPS with large- and small-scale slopes of $-2.7$ and $-3.7$, respectively, with the break point at 130 pc (details on how simulated images are produced are given in Section 3.3).  Our tests confirmed that the rolling SPS method successfully fits double power-law for the simulated image with two slopes, while clearly preferring a single power-law for the other two simulated images. 
As shown in Figure~\ref{fig:Rolling_SPS_Sim_fTest} in the Appendix,
the rolling SPS applied on the single-slope simulated images produced a mean $\gamma$ that underestimated (making the slope more steep) the input value by less than 10\%, with a standard deviation of $\sim0.2$.  The reason for this underestimation is the residual Gibbs phenomenon that corrupts both large and small spatial scales but has a stronger effect on large spatial scales due to its $\sin(k)/k$  behavior. The standard deviation of the recovered slope values is a function of kernel size as the Gibbs effects becomes smaller when we use larger sub-images. Increasing the  kernel  size past 1.5 kpc 
reduced the standard deviation of slope values and brought the mean closer to the input value 
(as shown in Figure~\ref{fig:meanSlopevsImageSize} in the Appendix).  Ultimately 1.5 kpc frames were deemed to produce acceptable results, as increasing the sub-image size decreases the amount of usable pixels at the edge of the image and conveys less information about small-scale structural variations.
Applying the rolling SPS to the simulated broken power law image, mean small- and large-scale slopes in the image again slightly underestimated the input values, and the mean calculated break point physical scale.

As a conclusion, our rolling SPS with the F-test distinguishes between single vs double power-law images. However, due to the residual Gibbs effect it recovers slightly steeper slopes. As shown in our tests in the Appendix, this introduces a systematic uncertainty of $\sim0.2$ for our selected kernel size. However, this underestimate is still within 1-$\sigma$ of our SPS fitting uncertainty.

\subsection{Structure Function}

In addition to the spatial power spectrum we also use the structure function to investigate turbulent properties of 2-dimensional images (either brightness temperature or column density). 
At the start, we integrate the image over all velocity channels and then normalize the integrated image by its maximum pixel value. Since the structure function is calculated in the image domain, it does not suffer from edge effects, commonly seen when Fourier transforming an image. In addition, the image can have any size and shape. However, one disadvantage of the structure function is that it covers a smaller range of spatial scales than the SPS.

We use Equation~\ref{e:SF} to calculate the structure function at each discrete pixel separation $r$, where $r'$ represents any arbitrary pixel in the image and $I$ represents either the normalized intensity or normalized column density of the pixel. 
%
For a given pixel separation $r$, there is a limited number of possible ways (or realizations) of selecting pixel pairs that can be calculated into the average squared difference of pixel intensities.
We calculate SF $SF_{r}$ for all realizations, and this allows us to estimate the mean $SF_{r}$ and the standard deviation over different realizations. 
Next we bin $r$ values by 0.05 pixels in log space. In doing so we use estimated uncertainties for each $r$ to calculate the weighted average and standard deviation of a given bin. The bar $\Delta v$ shows that the structure function is calculated for a specific thickness of velocity channels. Throughout most of the paper we work with the column density images and $\Delta v$ is the full velocity range of a given spectral line data cube (129 km/s in the case of the SMC ATCA + Parkes data cube, or 198 km/s in the case of the LMC ATCA + Parkes data cube). However, in Section 4.4 we investigate individual velocity channels and there $\Delta v=1.65$ km/s. 
\begin{equation}
    SF_{r}=\langle[I(r')-I(r'+r)]^2\rangle_{r} |_{\Delta v} .
    \label{e:SF}
\end{equation}

As with the SPS, we compute the rolling SF by calculating the SF slope for a small kernel and running the kernel across the whole image.

If the SPS of a 2D distribution is a power-law function with a spectral index $\gamma$ (such that $P(k) \propto k^{-\gamma}$), then the structure function will also be a power law with an index $\alpha$, such that $SF(r) \propto r^{-\alpha}$. If $2<\gamma<4$, then the two indices are related by the equation \citep{Simonetti84}:

\begin{equation}
  \alpha = \gamma - 2 .
\end{equation}
However, the above relation between the SF and SPS, in practice, is an approximation. As explained by 
\cite{Hou98}, due to the limited image size and angular resolution, the SF can deviate from a simple power-law on either small or larger spatial scale, complicating the selection for the range of spatial scales used to derive the slope. The level of deviation is a function of image size, resolution and $\gamma$ value, and is difficult to predict analytically.

\begin{figure}[h!]
  \centering
  \includegraphics[width=0.5\textwidth]{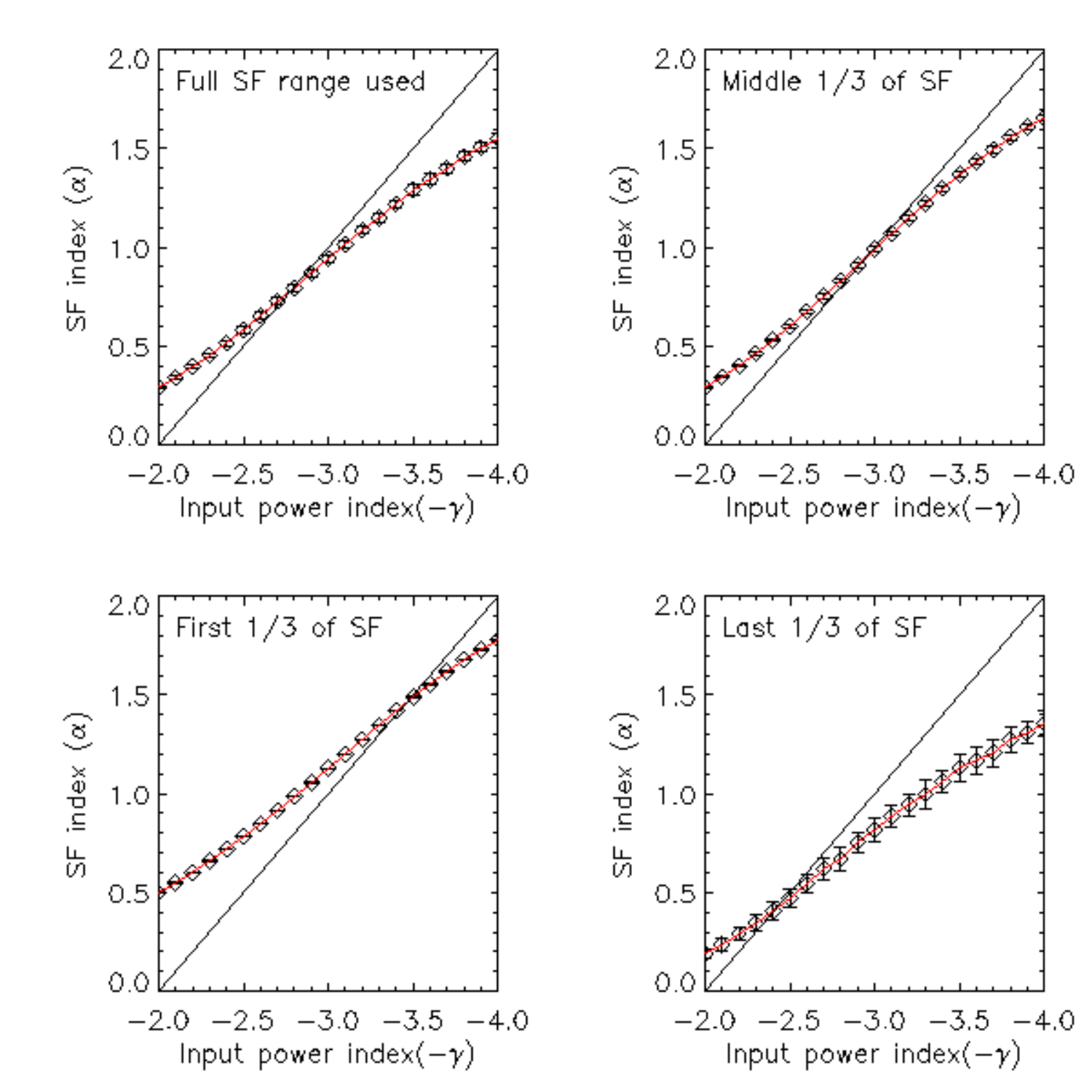}
  \caption{Comparison between the measured slopes of the structure function (SF) versus the input spatial power spectrum slope. We generated 100 fBm images with the input SPS slope from $-2$ to $-4$ at 0.1 intervals, all images have a size of 255 by 255 pixels.  The mean SF slope and standard deviation estimated for simulated images are plotted when the SF slope was calculated over the entire SF range (top left), the first 1/3 (bottom left), the middle 1/3 (top right), and the last 1/3 (bottom right) of the SF.  The derived SF slope deviates from the  true input SPS value.  For steep SPS slope ($-3.5$) measuring the first 1/3 of the SF range will produce the closest results, conversely if the SPS is shallow ($-2.5$) then the last 1/3 is a better range for measuring the SF slope.  For SPS slopes in the range $-2.7$ to $-3.5$, the  range using the central portion of the SF to calculate the slope is reasonable (within a few tenths of the actual SPS slope). 
  }
  \label{fig:SF-PS-simulations-2}
\end{figure}

Figure~\ref{fig:SF-PS-simulations-2}
demonstrates the effect of divergence between the SPS and SF slopes.
We use simulated images of 2D fractal Brownian motion (Section 3.3) to explore how the input power law slope affects the shape of the structure function to constrain the best spatial range for measuring the structure function slope.   We generated 100 simulated fBm images for SPS slope $\gamma=-2$ to $-4$ at 0.1 intervals, and for the image size of 255 $\times$ 255. 
In Figure~\ref{fig:SF-PS-simulations-2} we show the relation between the true (input) 
SPS slope $\gamma$ and the estimated SF slope $(\alpha$). The top right panel shows the relation when the full range of data points is used for fitting $\alpha$, while in the subsequent panels we use the middle 1/3, the first 1/3, and the last 1/3 of the spatial range for fitting, respectively. These figures show how much the measured $\alpha$ deviates from the relation given by Equation (6). For SPS slopes in the range $-2.7$ to $-3.5$, the range of slopes that is of interest when studying the SMC and LMC, the  middle 1/3 of the spatial range results in the SF slope that is within a few tenths of the actual slope expected based on the SPS slope. 
Based on these tests with simulated images,
we conclude that when estimating the SF slope
 using the middle 1/3 of the spatial fitting range is the most reasonable, and accurate within $\pm 0.2$, for
 the range of SPS slope from $-2.2$ to $-3.8$ (this range is of interest in our study based on previous SMC and LMC statistical investigations). Based on this, for our SF measurements in this paper we will always fit using the middle third of the range of spatial scales provided. 

\begin{figure}[h!]
  \centering
  \includegraphics[width=0.5\textwidth]{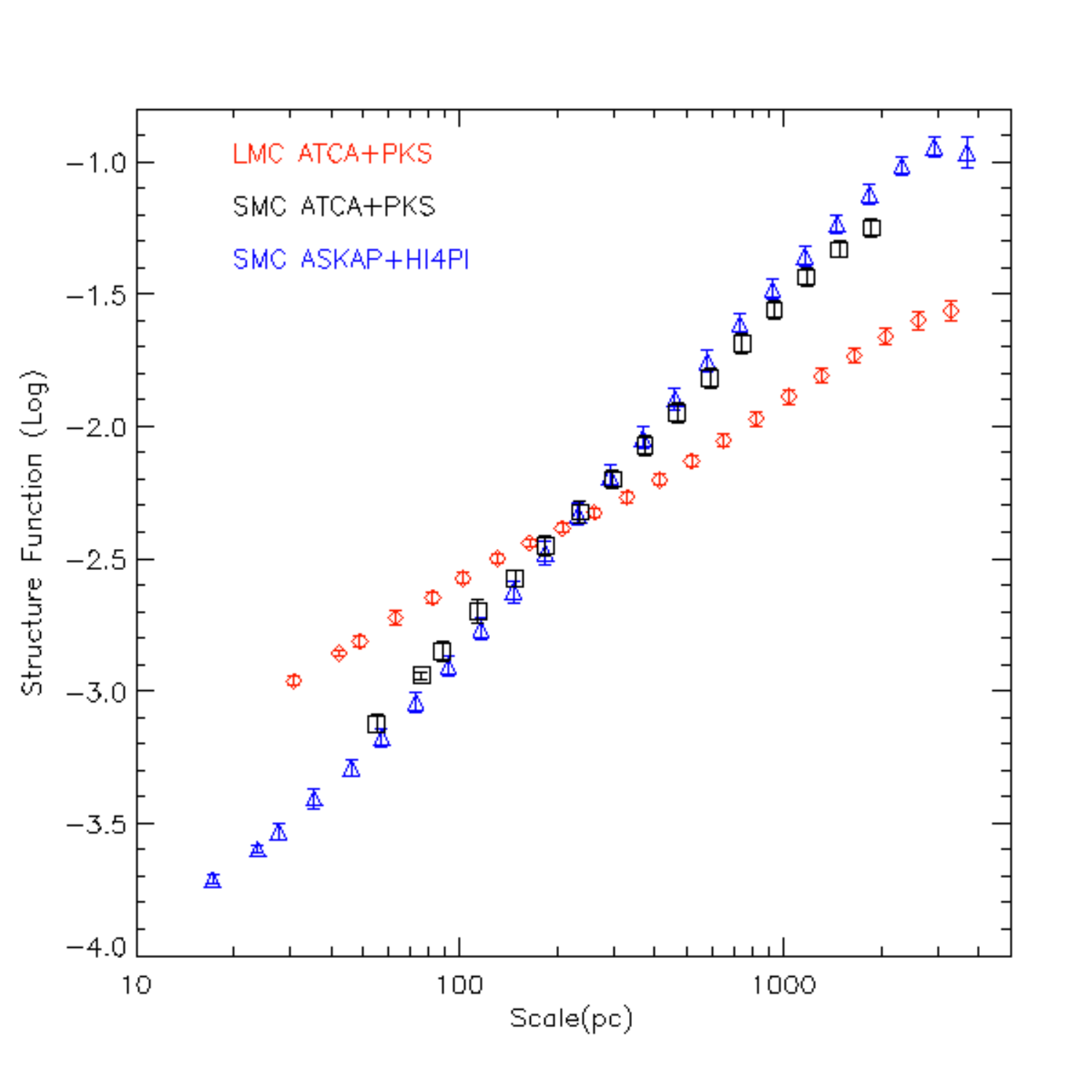}
  \caption{Global HI structure functions of the entire LMC and SMC HI integrated intensity images.}
   \label{fig:LMC-SMC-global-SF}
\end{figure}

\subsection{Fractal Simulations}
\label{s:fractal-sims}

We performed simple 2D fractal Brownian motion (fBm) simulations
to test the reliability of our measured SPS and SF 
slopes, as well as to estimate the degree of their
statistical fluctuations. 
%
To generate simple (without noise) 2D fBm images ($F$),
we use the prescription from 
\cite{Miville-Deschenes03}.
We
briefly summarize here
their approach.  To create a fBm simulated image whose spatial power spectrum 
follows a perfect power law (with random phase) we first start by generating a power-law isotropic amplitude (their equation 3):
\begin{equation}
  A(\mathbf k) = A_0k^{\gamma/2} .
\end{equation}
Here, \textbf {k} is a two dimensional wavevector, whose length is the wavenumber $k$.
$A_0$ is the normalization value and $\gamma$ is the SPS slope.  The phase $\phi$(\textbf k) is randomly generated between -$\pi$ and $\pi$, additionally we constrain $\phi$(\textbf{-k}) = -$\phi$(\textbf k) to ensure the image in direct space is real.  We now create the real and imaginary components of the Fourier transform:
\begin{equation}
  Re[\mathit{F}(\mathbf {k})] = A(\mathbf {k}) \cos[\phi (\mathbf {k})],
\end{equation}

\begin{equation}
  Im[\mathit{F}(\mathbf {k})] = A(\mathbf {k}) \sin[\phi (\mathbf {k})],
\end{equation}
The Gaussian random field $F$ (fBm) is the inverse Fourier transform of \textit{F}(\textbf {k}).
Using this technique we construct simulated images of various sizes and with the varying SPS slope.

\section{SMC Structure Function (SF) and Spatial Power Spectrum (SPS) Analysis}
\label{s:SMC}

\subsection{Global Properties}

Figure~\ref{f:SMC-ASKAP-img} shows the HI integrated intensity image of the SMC obtained using ASKAP and HI4PI data.
Although obtained with only 16 ASKAP antennas, this image already has a factor of three higher angular resolution than the previous ATCA observations. In terms of sensitivity, the ATCA+Parkes dataset is a factor of two more sensitive. The ASKAP image also has a slightly larger extent relative to the previous ATCA+Parkes image. The corresponding peak HI column density based on this image is in excellent agreement with that from Stanimirovic et al. (1999).

Figure~\ref{fig:LMC-SMC-global-SF} shows global SFs calculated by using the whole SMC and LMC HI column density images. For the SMC we show both the old ATCA + Parkes and the new ASKAP + HI4PI data sets to demonstrate how new HI observations have both higher spatial resolution and a slightly larger spatial coverage. 
The two SMC SFs agree well throughout most of the spatial range, with some small departures at the small and large ends. These SFs also show slight turnover at the largest spatial scales due the effects we discussed in Section 3.2. 
In addition, this figure clearly shows that the SF slope of the LMC HI is significantly more shallow than the slope of the SMC. 
Finally, while the SMC SF is linear (in the log-log space), the LMC SF shows a break around 100-200 pc  (we discuss this in Section~\ref{s:LMC}). The break in the SPS of the LMC has been found to be similar to the estimated HI scale height of $\sim180$ pc 
(Kim et al. 1998).
The lack of break in the SPS of the SMC can be explained as being due to the lack of a ``thin disk''. Indeed, \cite{Stanimirovic04} provided a rough estimate of the HI scale height of 1 kpc based on the typical size of large HI shells as well as the velocity dispersion.

\subsection{Spatial Distribution of the SMC HI SF and SPS Slopes}

We next run a circular kernel with a diameter of 51 pixels 
(corresponding to $\sim24'$ and covering the spatial range 8-415 pc) 
across the SMC HI integrated intensity image from ASKAP + HI4PI and in each kernel calculate the structure function. For each structure function we fit a single power-law function (based on previous studies) and map out the estimated slope spatially across the SMC. 

\begin{figure}[h!]
  \centering
 \includegraphics[width=0.5\textwidth]{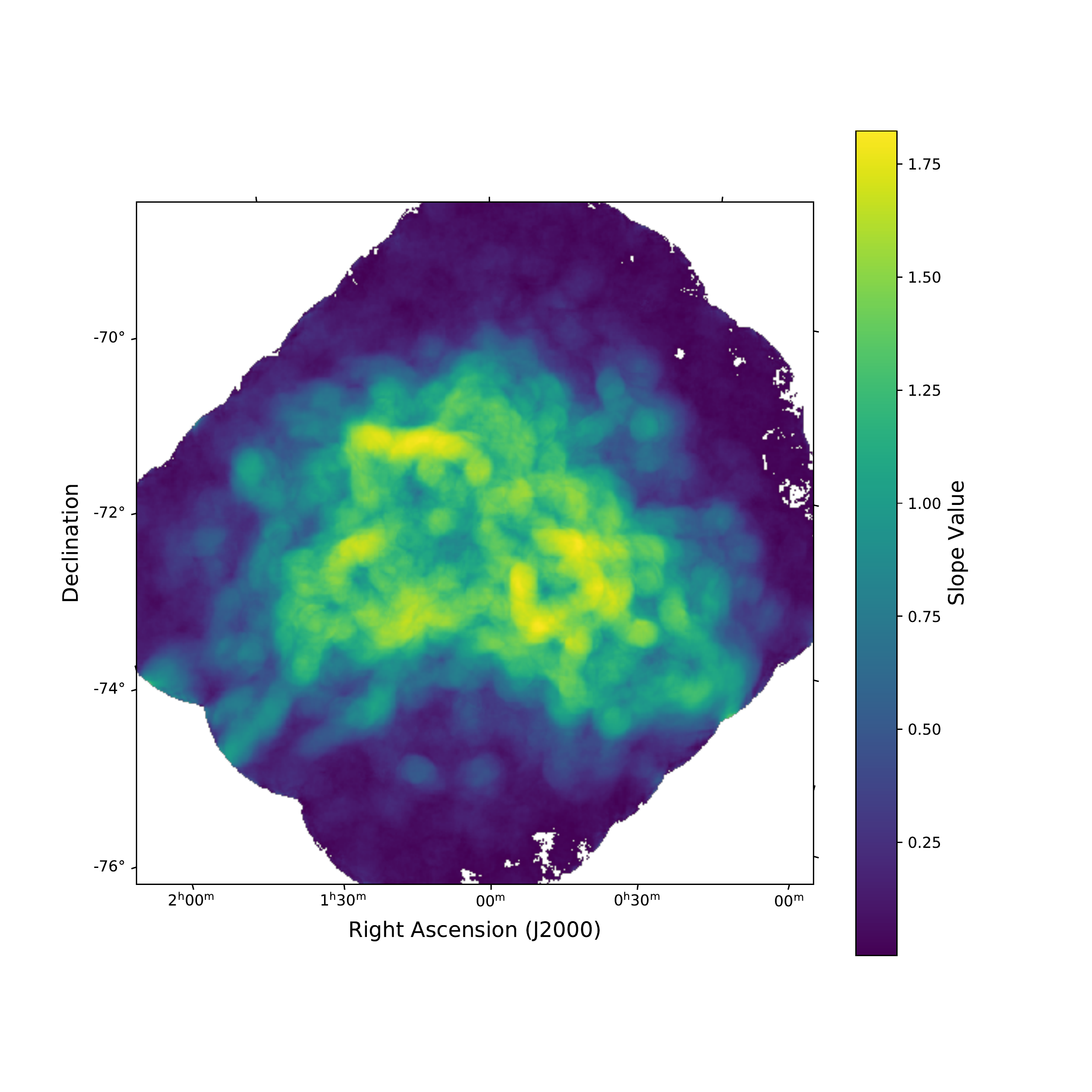}
   \caption{The SMC structure function slope image. We used a circular kernel of 51 pixels in diameter (corresponding to 24$'$)
   and applied the rolling SF calculation on the HI ASKAP+HI4PI integrated intensity image of the SMC. 
    A small fraction of pixels with a slope value below 0, located in the far image outskirts which are noise dominated, have been removed.
   Motivated by previous studies, e.g. Stanimirovic et al. (1999), 
   a single power-law function was fit in each kernel. The effective angular resolution of this image is 24$'$. Based on our analysis with simulated data, a typical uncertainty of the SF slope is $\sim0.2$. While this image shows small spatial variations, these are consistent with statistical fluctuations and do not stand out as being statistically significant.}
    \label{fig:SMC-SF-slope}
\end{figure}

\begin{figure}[h!]
  \centering
   \includegraphics[width=0.5\textwidth]{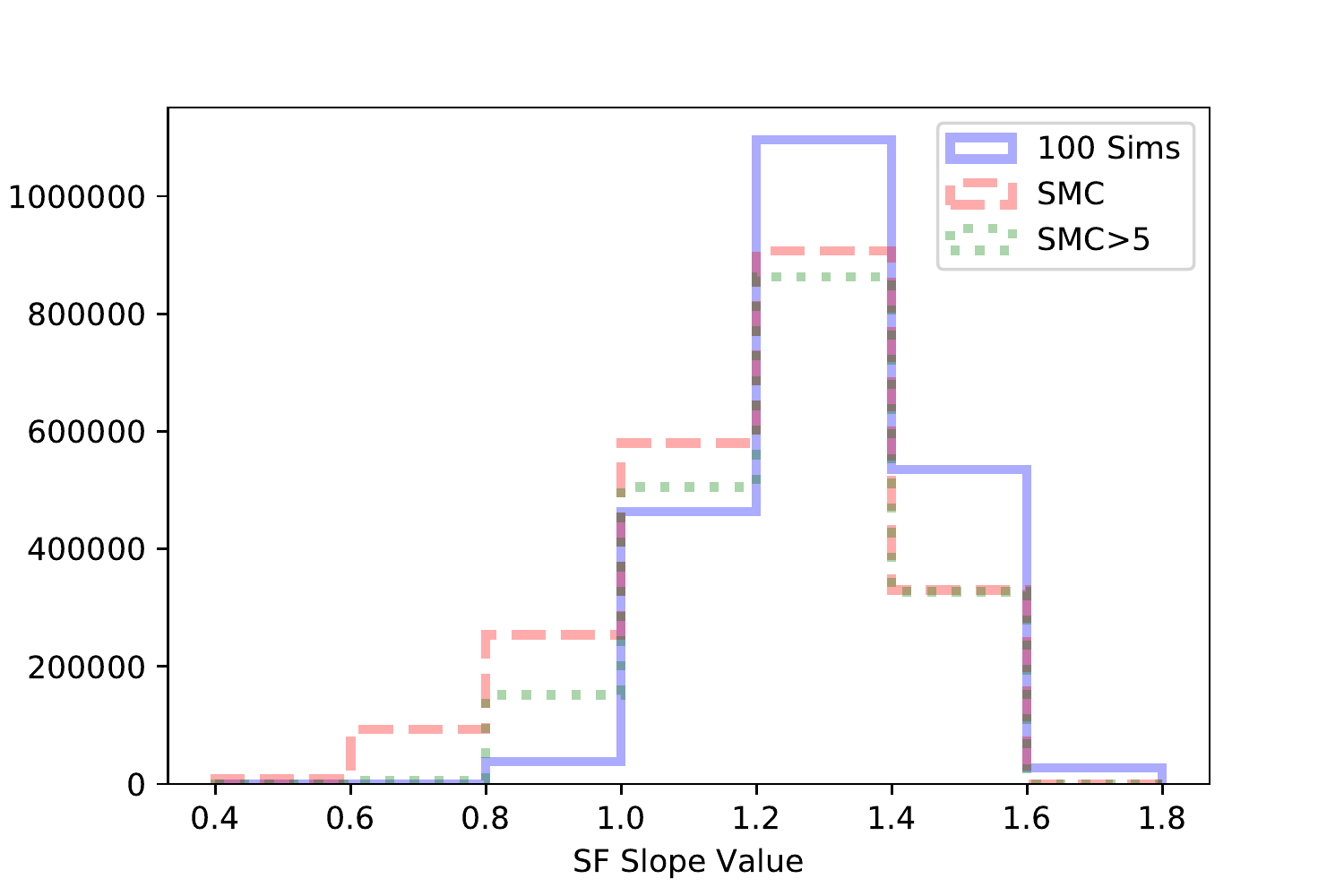}
   \caption{The histogram of the SMC SF slope (with a bin spacing of 0.2) showing data from Figure~\ref{fig:SMC-SF-slope} (red) and after excluding pixels with S/N$<5$ (green). The histogram of SF slopes for the 100 simulated images, generated to have a SPS slope of $-3.4$, is shown in blue.}
   \label{f:histogram-SMC-slope}
\end{figure}

The resultant image of the SF slope is shown in Figure~\ref{fig:SMC-SF-slope} and has a resolution of $24'$. 
The SF slope ranges between $\sim1.0$ and 1.6.
Using our arguments in Section 3.2, this range corresponds to the SPS slope $-3.0$ to $-3.6$. While the image shows small spatial variations, these are not significant and are consistent with statistical fluctuations as shown in Figure~\ref{f:histogram-SMC-slope}. 
In Figure~\ref{f:histogram-SMC-slope} we show the histogram of SF slopes for the SMC image and 100 simulated images. 
We generated 100 fBm images all having the input SPS slope of $-3.4$. We then treated each simulated image like observations and calculated the rolling SF slope. The histogram of all SF slopes is shown in solid blue.
When considering SMC S/N$>5$ pixels (dotted green line), essentially the simulated SF slope is in excellent agreement with observations.
This demonstrates that for the same SPS slope statistical fluctuations will result in a small range of structure functions -- this range is consistent with what we find across the SMC. This exercise also shows that a typical uncertainty of our calculated SF slopes is $\pm0.2$ (estimated as the standard deviation of the simulated distribution), in agreement with our considerations in Section 3.2.

\begin{figure}[h!]
  \centering
 \includegraphics[width=0.5\textwidth]{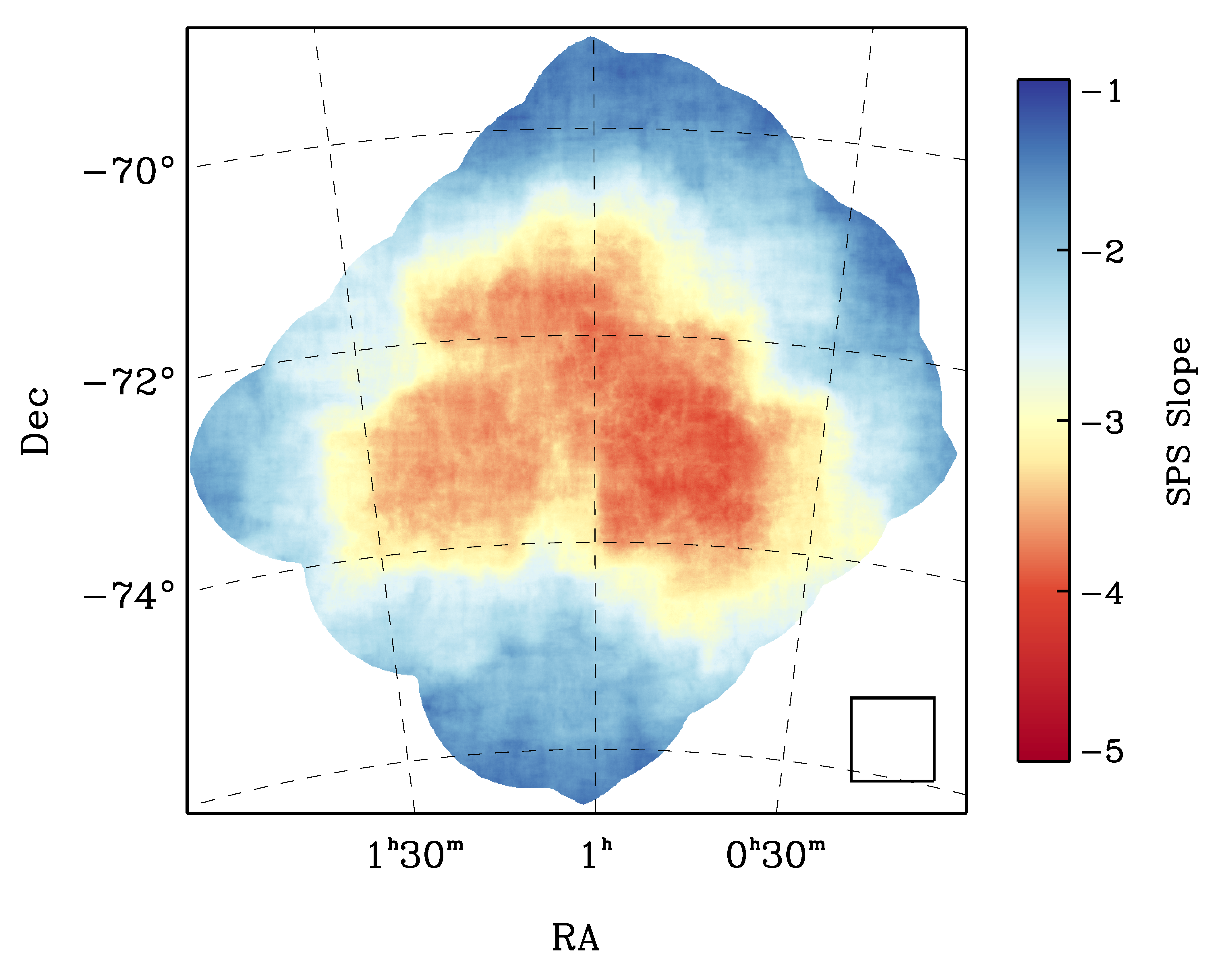}
   \caption{The SPS slope image calculated from the SMC's HI (ASKAP+HI4PI) integrated intensity image.  Each pixel represents the center of an 850 pc square kernel, shown in the lower right corner, in which SPS was calculated. A single power-law function was fit in each kernel. The effective angular resolution of this image is $\sim49'$. A typical uncertainty of the SPS slope per kernel is 0.1-0.3.}
   \label{fig:SMC-SPS-slope}
\end{figure}

We get a very similar result if we apply the rolling SPS and calculate the spatial distribution of the SPS slope, shown in Figure~\ref{fig:SMC-SPS-slope}.   To compute this image we use a kernel size of 850 pc and shift by 8 pc (the pixel size of the HI integrated intensity image).
We fit a single slope based on previous studies, and also after investigating SFs of various sub-regions. In addition,  we also allowed for fitting a double power law and used the F-test to decide which of the two models fit the data better. We found that for some pixels the $F$-test preferred a double power law. Despite this, the probability distributions of single, large-scale, and small-scale slopes across the galaxy all peak near the same value, a characteristic also present in single-slope simulated images.  
The rolling SPS image (shown in Figure 6) has a median slope of $-3.0 \pm 0.5$ (error estimated by MAD) and a mean slope of $-3.0 \pm 0.6$ within the region studied by Stanimirovic \& Lazarian (2001).  These values are in agreement with their SPS slope calculation of $-3.3 \pm 0.01$ using the entire HI column density image. The median fitting uncertainty across all kernels is 0.15.


The SF and SPS slope images suggest highly uniform turbulent properties for the HI in the SMC. This is a surprising result considering the SMC-LMC-MW interactions
where being the lowest mass the SMC has suffered significantly and a lot of the SMC HI has been pulled out to form the Bridge and the Stream.
Our estimated slope values, and their range, are in agreement with previous studies, e.g. Stanimirovic et al. (1999), Stanimirovic \& Lazarian (2001) and Nestingen-Palm et al. (2017). 


\subsection{Correlations with the SFR and stellar surface densities}

\begin{figure}[h!]
  \centering
   \includegraphics[width=0.5\textwidth]{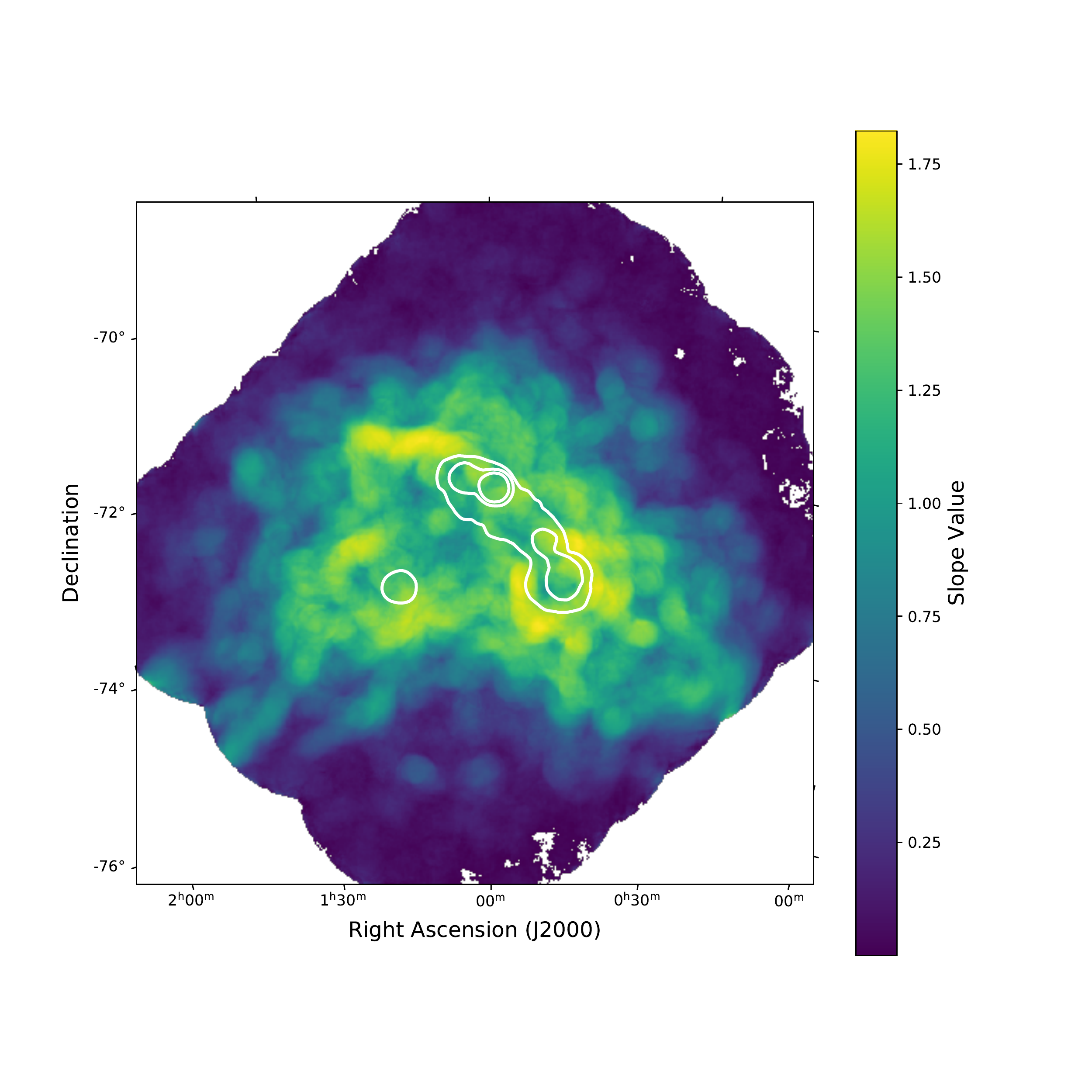}
   \includegraphics[width=0.5\textwidth]{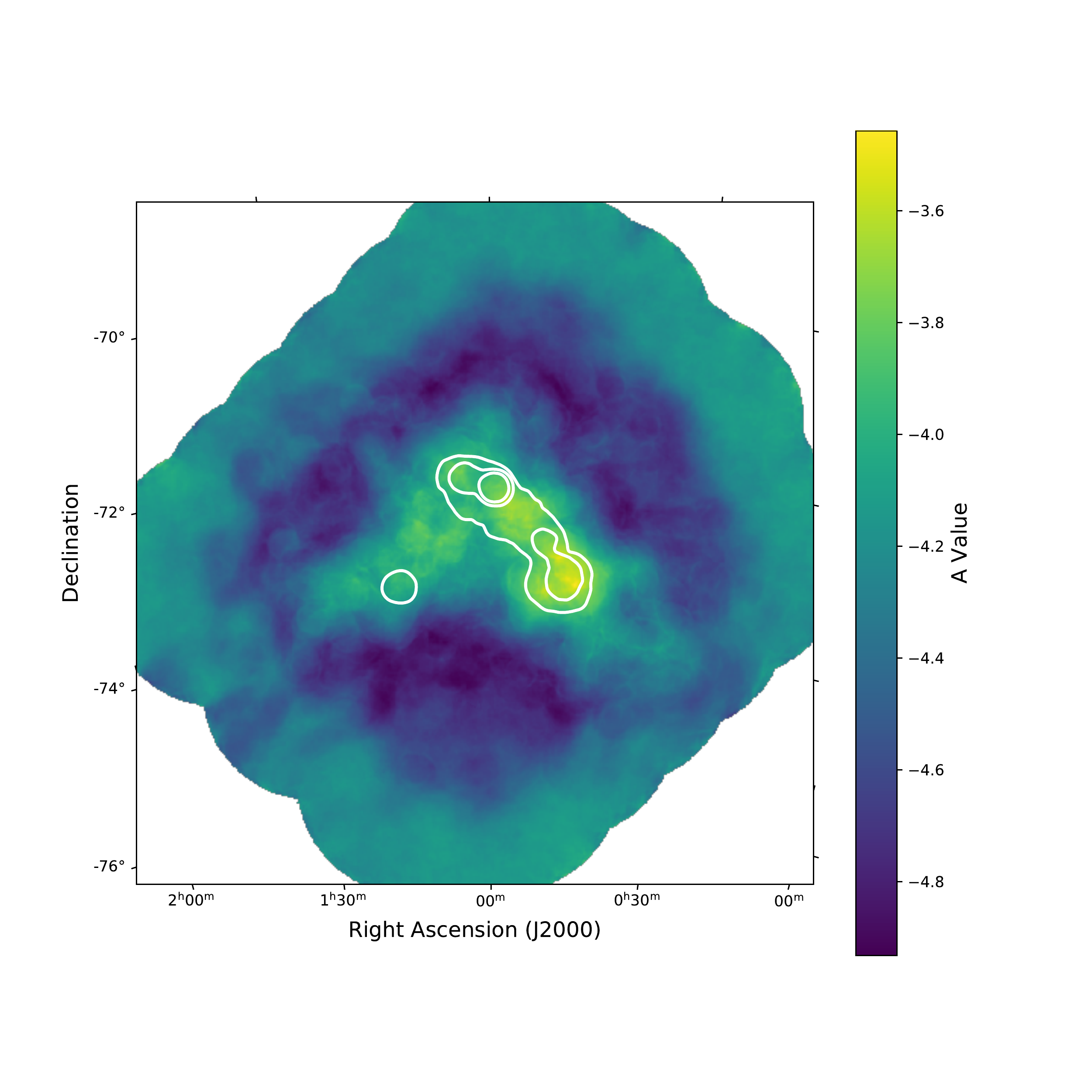}
  \caption{(Top) The rolling Structure Function slope of the  SMC's HI integrated intensity image obtained from the ASKAP+HI4PI data,  same as in Figure 4. 
   The SFR surface density is overlaid with contours at 0.005, 0.012, and 0.024 MJy/sr. 
   (Bottom) The normalization value of the fitted SF power-law functions overlaid with the SFR surface density.}
   \label{fig:SFslopevsSFR}
\end{figure}

Figure~\ref{fig:SFslopevsSFR} shows the slope and the normalization of the rolling SF calculated using the SMC HI integrated intensity image, overlaid with the SFR surface density contours (smoothed to the same resolution). Numerical simulations (e.g. Walker et al. 2014, Grisdale et al. 2017) suggest that stellar feedback steepens the SPS or SF slope by destroying clouds and shifting the power from small to large spatial scales. We would therefore expect to see a larger SF slope at the position of sites of recent star formation. However, in the SMC we do not find any obvious correlation between the SFR surface density and the SF slope or the normalization value.
Similarly, we do not find that distributions of either the SF slope or the normalization value correlate with the stellar surface density.


\subsection{Probing velocity fluctuations}

Numerical simulations of stellar winds by \cite{Offner15} have suggested that stellar winds can significantly influence the
velocity power spectrum, while the density power spectrum can be less affected.
We have so far considered only fluctuations in the HI integrated intensity images.
 Lazarian \& Pogosyan (2000) suggested that the SPS of integrated intensity images
probes density fluctuations, while
velocity fluctuations are more prominent when examining intensity fluctuations of individual velocity channels.
In addition, as velocity fluctuations can contribute to intensity fluctuations, the SPS slope of individual velocity channels is
expected to be more shallow than the slope of the integrated intensity image.
Stanimirovic \& Lazarian (2001) investigated the SPS of the entire SMC ATCA + Parkes data set and showed that the SPS slope
of the entire SMC calculated for a velocity thickness of $\Delta v=1.65$ km/s is indeed more shallow than the slope of the entire integrated intensity image (2.8 vs 3.3). 
We now investigate spatial variability of the SF slope at $\Delta v=1.65$ km/s.

\begin{figure}[h!]
  \centering
   \includegraphics[width=0.5\textwidth]{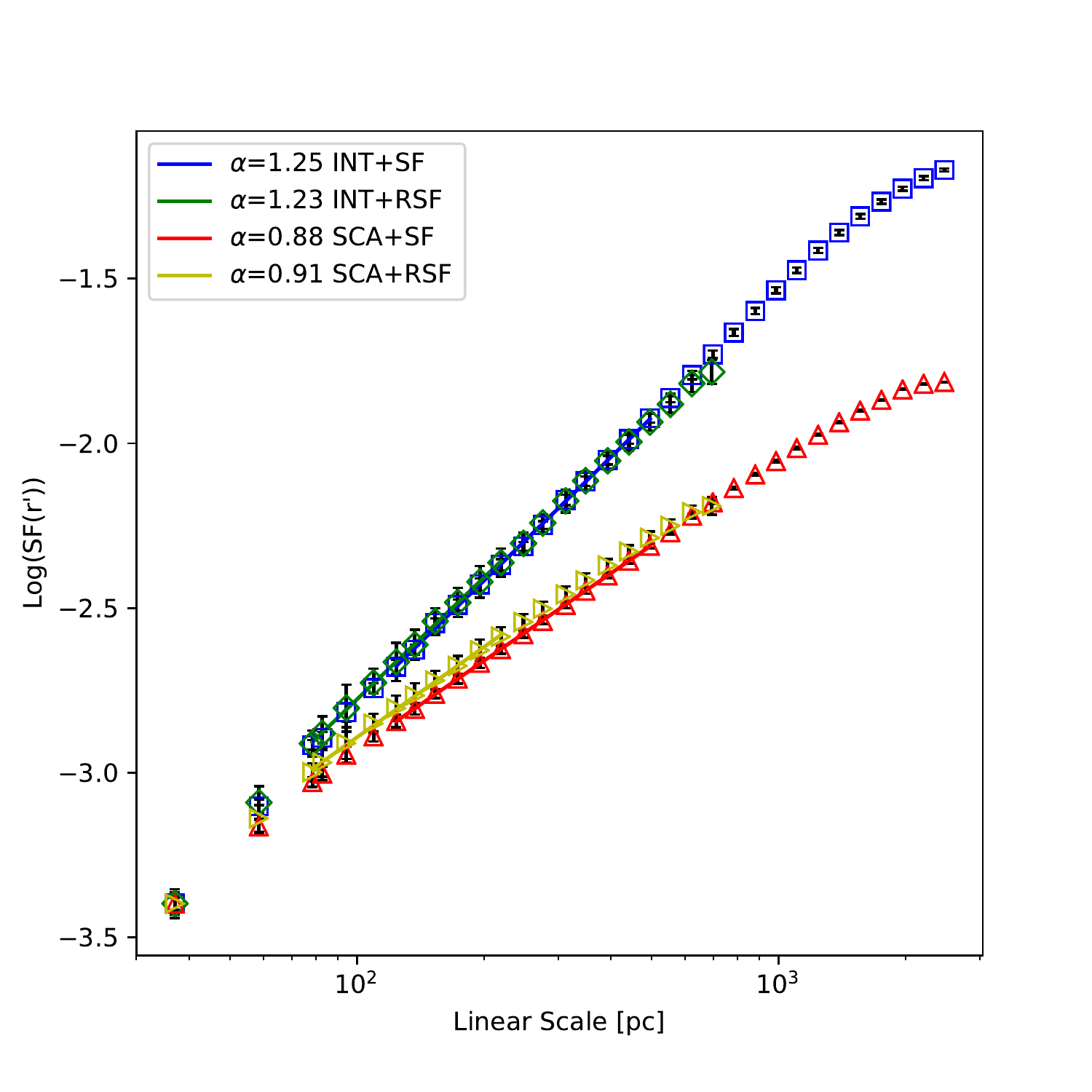}
   \caption{SMC Structure functions: SF of the HI velocity-integrated intensity image (squares), rolling SF average of the HI velocity-integrated intensity image (diamonds), average SF of the individual velocity channels (110 to 210 km/s range, triangles), and an average of the rolling SF averages of the individual velocity channels (right-hand triangle). Note, we have normalized all SFs at $\sim25$ pc by subtracting all bins by the difference of their first bin.}
   \label{f:full-channel-SF}
\end{figure}

For each individual velocity channel in the emission-dominated velocity range 110-210 km/s (60 channels in total), 
we calculate the SF. We can do this using two different approaches.
First, for each one of 60 velocity channels we calculate the SF using equation (5), 
then average all 60 SFs. The result is shown in Figure~\ref{f:full-channel-SF}  using red triangles.
Second, for each individual velocity channel we run the rolling SF method and calculate the SF in each 51-pixel wide kernel (this results in 776,520 kernels x 60 channels = $4.7\times 10^7$ individual SFs),
then average all SFs. The resultant SF is
shown with yellow right-hand triangles in Figure~\ref{f:full-channel-SF}. These two SFs are in excellent agreement for the overlapping range of spatial scales.

As a comparison, in the same figure we show SFs for the velocity-integrated HI image (with $\Delta v=129$ km/s).
Again, we can calculate this using the two different methods.
First, the blue squares show the SF calculated using the whole SMC {\bf velocity-integrated} intensity HI image, this is the same SF as the one shown in Figure 3 (blue triangles). 
Second, using the rolling SF method we can calculate the SF for 776,520 kernels in the HI integrated intensity image. The average of these SFs is shown with green diamonds. 
Again, the two SFs are in excellent agreement over the overlapping range of scales.

However, it is obvious that the SFs of the integrated intensity image have steeper slope than SFs of individual velocity channels, in agreement with expectations from Lazarian \& Pogosyan (2000). This result agrees with previous studies, e.g. Stanimirovic \& Lazarian (2000) and Nestingen-Palm et al. (2017). While this is not a new result, we have shown that calculating rolling SFs produces consistent results (when compared over the same range of spatial scales) to more traditional approaches of using an entire image.
 We have also compared the SF slope images calculated for individual velocity channels with the SFR surface density but did not find any obvious correlations.


\section{LMC Statistical Analysis}
\label{s:LMC}

\subsection{Global Properties}

\begin{figure}[h!]
  \centering
  \includegraphics[width=0.4\textwidth]{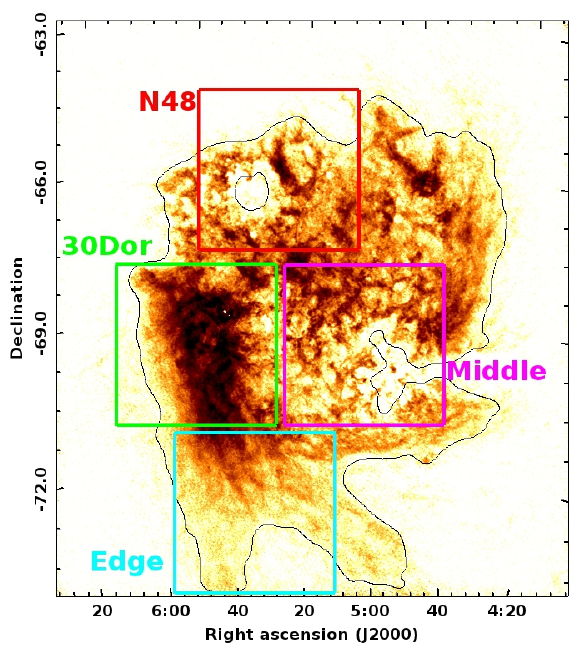}
  \includegraphics[width=0.4\textwidth]{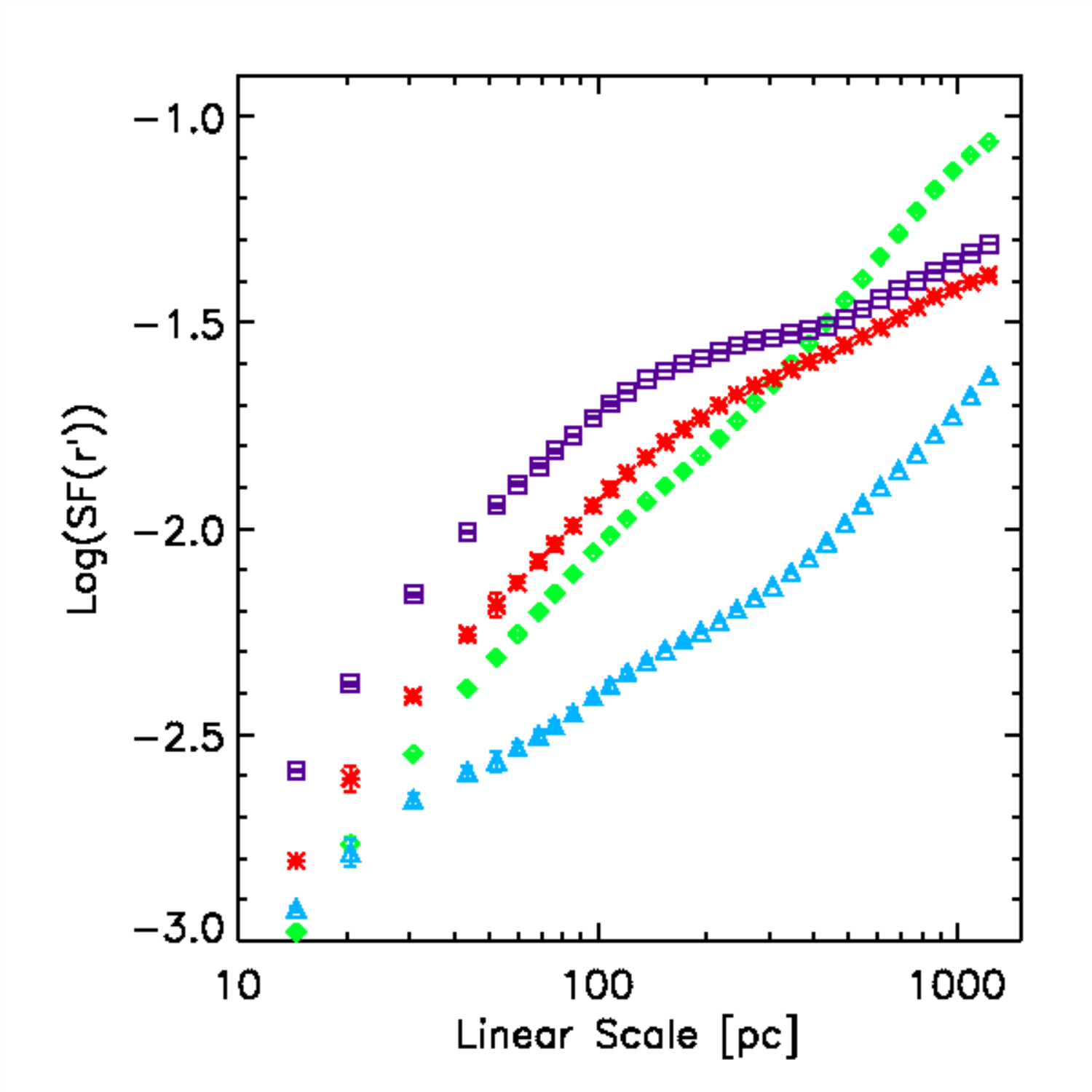}
  \caption{(top) The HI integrated intensity image of the LMC  with 4 selected regions:
  a region around the star-forming region N48 (red); 30 Doradus (green); roughly the middle area of the LMC (magenta) and the bottom edge of the LMC (cyan).  All areas are square and 2.8 kpc across. The black contour line shows where the intensity is 3-$\sigma$ above the noise level. (Bottom) SFs of the 4 LMC regions (red asterisks, green diamonds, magenta squares, cyan triangles). Uncertainties are typically smaller than the symbol sizes.}
  \label{fig:LMC-subregions}
\end{figure}

As shown in Figure~\ref{fig:LMC-SMC-global-SF} the SF slope of the LMC HI  is more shallow than the slope of the SMC. In addition, while the SMC SF is linear (in the log-log space), the LMC SF has a likely break around 100-200 pc which is even more pronounced when examining SFs of more localized regions.  In Figure~\ref{fig:LMC-subregions} (left) we show 4 selected regions which display significant variability of the SFs within the LMC. 
The SF for those 4 regions are shown in Figure~\ref{fig:LMC-subregions} (right). The 4 areas include 30 Dor, center of the LMC, and two positions in the outskirts of the LMC. This figure shows not only that SFs change significantly across the LMC, but that likely multiple breaks exist.


The influence of the LMC disk on the spatial power spectrum has been noticed in several studies. Besides Elemegreen et al. (2001) and Padoan et al. (2001), more recently Block et al. (2010) used Spitzer IR images at several wavelengths 24, 70 and 100 $\mu$m and showed a clear break at a spatial scale of 100-200 pc. As we discussed in Section 3.2, structure function suffers at both small and large spatial scales due to the limited image resolution and size. Adding further complexities caused by the presence of a prominent disk-like structure would introduce further deviations and make interpretation of SFs slopes and possible breaks even more complex. Because of this, when investigating spatial variations of turbulent properties across the LMC we use only the SPS. 

\subsection{Spatial Distribution of the LMC HI SPS Slope}

\begin{figure*}[h!]
  \centering 
 \includegraphics[width=\textwidth]{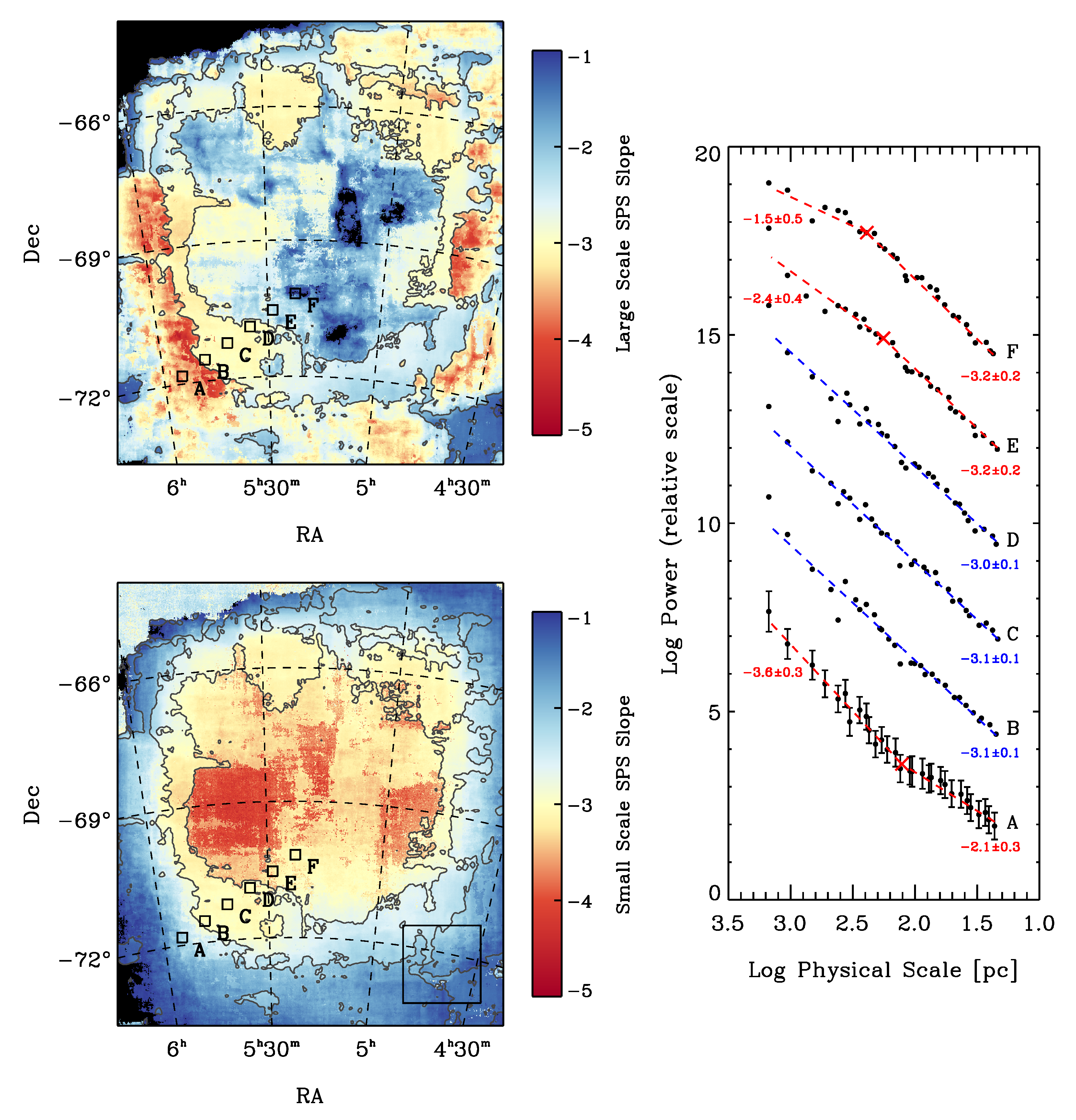}
 \caption{(Top-left) Large-scale SPS slope for the LMC HI column density image.  Each pixel represents the center of a 1500x1500 pc kernel in which the slope was calculated.  Contours enclose regions in which a single power-law model better fits the data at a 5\% significance level, and where slopes from this simpler model are plotted.  Black pixels indicate fitted slopes outside the colorbar range.  (Bottom-left) Small-scale SPS slope across the LMC, including a box in the bottom right corner indicating kernel size. The median fitting uncertainty for estimated slopes on both small and large scales is of order 0.2-0.4. (Right) Relative SPS averaged over 121 kernels centered within the squares shown on the images.  SPSa are fit with a broken power law where applicable, shown as  red dashed lines, and with a single power law shown in blue where the $F$-test signals a failure to reject the simpler model.  Slope values and one sigma uncertainties are displayed at corresponding ends of the spectra along with red crosses at the detected break points.  Error bars are identical in each spectrum as prescribed in Section 3.1.}
   \label{fig:LMC_gamma_large}
\end{figure*}

\begin{figure}[h!]
  \centering
 \includegraphics[width=0.5\textwidth]{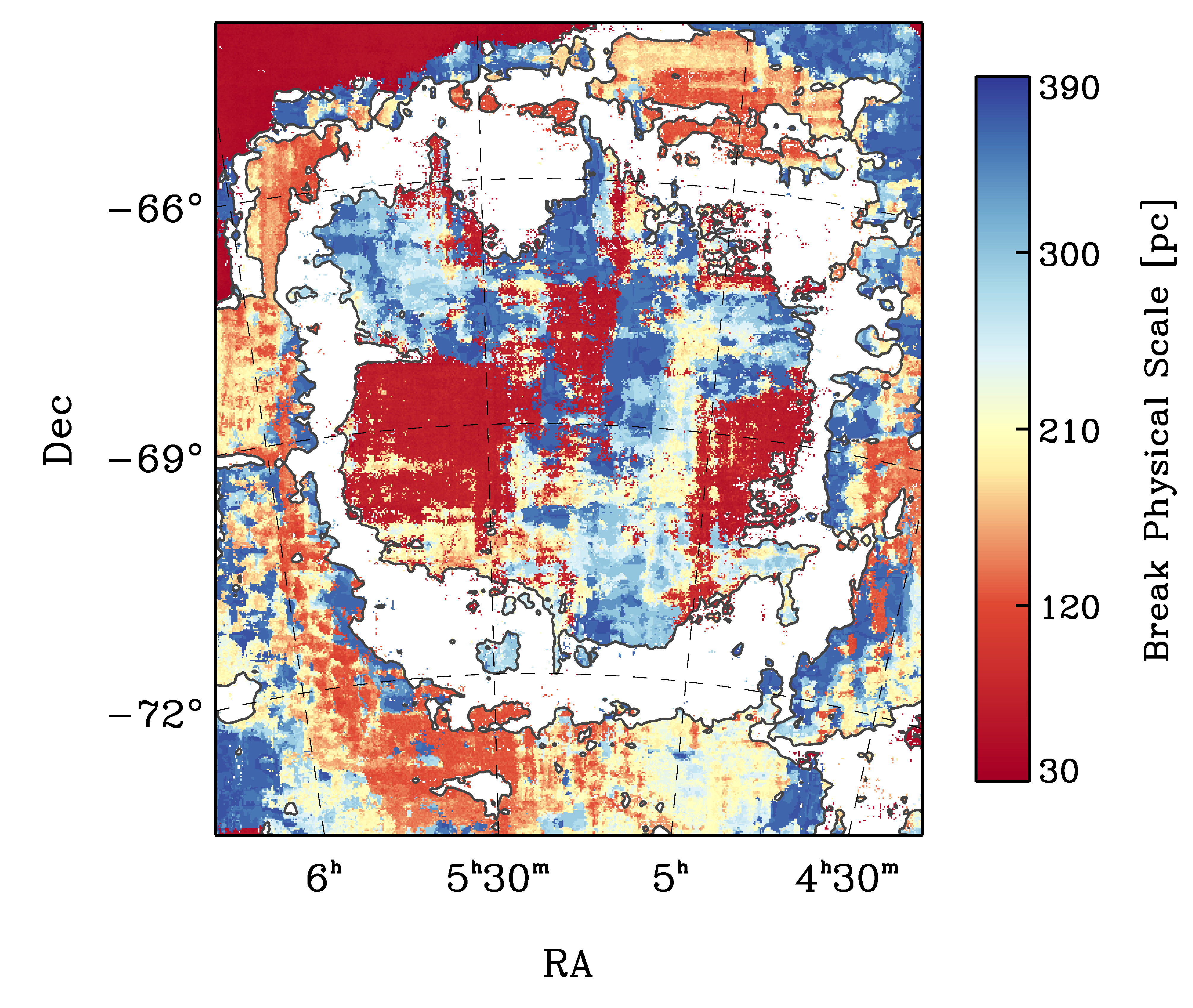}
   \caption{Physical scale corresponding to the break point in the LMC SPS.  The rolling SPS samples scales from 15-1500 pc, but 15\% of the bins are discarded at each extreme, leaving potential break points between 34 and 415 pc. Black contours enclose white pixels where a single power-law model better fits the data at a 5\% significance level.}
   \label{fig:LMC_dist}
\end{figure}

We apply the rolling SPS method on the HI integrated intensity image of the LMC.
For each kernel where we calculate the SPS, we test whether a single or a broken power law fits the SPS better, in a statistically significant manner, by employing  the $F$-test (see Section 3.1.1).
The large- and small-scale slope distributions across the LMC are shown in Figure~\ref{fig:LMC_gamma_large}, and the corresponding physical scales of the break points are plotted in Figure~\ref{fig:LMC_dist}.  Contours enclose regions where we fail to reject the null hypothesis, i.e. where a single power-law model is a more appropriate fit.
The single power-law fit is preferred primarily 
in an annular region and for 30.0\% of pixels.
The median fitting uncertainty for estimated slopes is 0.2 for a single power-law fit, and 0.2-0.4 for the large- and small-scale slopes, respectively.
We also show in Figure~\ref{fig:LMC_gamma_large} several example spatial power spectra obtained at representative positions A-F. Each shown SPS was obtained by averaging SPS calculated for 121 individual kernels all centered near a particular position.

In agreement with previous studies, e.g. Elmegreen et al. (2001), Padoan et al. (2001), Block et al. (2010), we find that most of the LMC requires a broken power-law to fit the SPS.
For most of the LMC area, the large spatial scales have a more shallow SPS slope $>-3$ vs $<-3$ at small scales. This is in agreement with previous studies (e.g. Elmegreen et al. 2001, Block et al. 2010) where the central portion of the LMC HI distribution is dominated by the disk and 2D turbulence which results in a shallow SPS slope (difference from $\sim-3.7$ to $\sim-2.2$ ).

Interestingly, on small spatial scales we find localized regions with very steep slope ($<-4$). The most prominent is the 30Dor region, and other regions are close to well-known HII regions: N48, N11 and N79. This is in agreement with predictions from numerical simulations for the influence of the stellar feedback and localized steepening of the SPS slope 
\citep{Grisdale17,Grisdale19}.
Stellar feedback via supernovae and stellar winds removes gas from clouds, or totally destroys clouds, effectively shifting power from small to large spatial scales. 
Considering our typical fitting uncertainty (0.2-0.4),
the difference between very steep slopes ($\sim -3.9$) caused by stellar feedback relative to the surroundings (slope of $\sim-3$) is clearly distinguished.

While throughout most of the LMC on large-scales the SPS slope is $>-3$, at the extreme LMC outskirts the slope becomes steep reaching values $<-3.8$ in two extended elongated regions. Again, this change clearly stands out even after taking 
into account uncertainties associated with the rolling SPS derivation.
We are confident that these steep slope regions are not caused by artifacts or image edges as we find them only in the LMC. Any artifacts caused by the FFT or our kernel size should affect equally both the SMC and the LMC, since kernels in either analysis span the same number of image pixels. In addition, as shown in Figure~\ref{fig:LMC_gamma_large}  (right), the SPS slope change 
is gradual not random. Pointings F to A demonstrate how
the SPS changes its shape and slopes when going from the center to the outskirts. While at pointing A  the SPS has essentially a convex shape, by the pointing F this has flipped into a concave shape.

These elongated regions with very steep large-scale slope are at the edges, or just beyond the stellar disk. On the South-West side, the steep SPS slope is in the region where optical observations have discovered warping and flaring of the stellar disk (Olsen \& Salyk 2002, Choi et al. 2018).
The steepening of the SPS slope we find likely indicates the transition from 2D to 3D turbulence due the presence of the HI disk flaring and a rapid increase in the disk scale-height. While within the thin HI disk turbulence is dominated by 2D motions, within the HI flare the disk thickness increases resulting in 3D turbulent motions.

On the South-East side of the disk we see similar steepening in the region at the edge of the stellar disk and right at the position of the ``leading HI edge''.  
One possible explanation is that we are seeing the South-East flare of the HI disk, just like on the South-West side. While this has not been seen previously, it was predicted in Besla et al. (2012) that stellar and gaseous disks are both warped and flared. 
In addition to the extended region of steep SPS slope, there is an indication that at this side of the disk the SPS slope has a more fragmented and patchy distribution. This could possibly be caused by the ram pressure effects as the LMC disk interacts with the hot MW halo.

The image of the SPS break points shows interesting spatial variations. 
We emphasize that both low and high extremes of the break point range are not well constrained with our fitting. 
The localized areas around 30Dor, N48, N11 and N79 seem to indicate the lowest disk thickness, $<50$ pc. The central regions of the LMC have a slightly higher thickness $\sim200-300$ pc. The elongated regions in the outskirts with steep SPS slope have a thickness of $>150-400$ pc. Most previous studies of the LMC have found a break at about 200 pc, in agreement with the estimate of the HI disk thickness of about 180 pc (Kim et al. 1998).

\subsection{Correlations with the SFR and stellar surface densities}



Figure~\ref{fig:gamma_large_small_SFR_stellar} shows the large and small scale SPS slopes overlaid with the SFR surface density and the stellar surface density (provided by the 3.6 $\mu$m Spitzer image). The SFR and stellar surface densities were smoothed to the resolution of the SPS slope images. With our coarse resolution that the SPS slope images have we can see that the 30 Doradus region overlaps with the localized region where the small-scale SPS slope is very steep. Similarly, 
positions of N79 and N11 coincide with locations where the small-scale SPS slope is steep.
In particular, N79 has recently been classified as a super star cluster candidate and is known to contain powerful outflows and an HII region \citep{Nayak19}.

The elongated regions with a steep slope that appear in the large-scale SPS slope image are positioned right at the edges of the SFR surface density contours. On the North-East side (where the HI leading edge is) the elongated region of very steep slope coincides with the edge of the stellar disk, while on the South-West side the elongated region with the steep slope is located beyond the stellar disk. This is in agreement with the HI distribution and also 160 $\mu$m Spitzer image which traces large dust grains well mixed with HI. On the North-East side the stellar distribution extends all the way to edge of the leading edge (seen in HI and 160 $\mu$m), while on the South-West side gas and large dust grains have a more extended distribution relative to the stars. Generally, the stellar contours occupy the region of shallow large-scale SPS slope ($>-3$).

In summary, the small-scale SPS slope shows a clear correlation with several most prominent HII regions where, in agreement with what is expected due to stellar feedback where the SPS slope becomes very steep as supernovae, stellar winds and outflows disrupt and destroy interstellar clouds. The large-scale SPS slope image has the most shallow SPS slope roughly in agreement with the extent of the stellar disk. The two elongated regions in the disk outskirts with a very steep slope are both located just beyond the stellar and star-forming disk of the LMC.

\section{Discussion}
\label{s:discussion}

We developed methods for mapping spatial variations of the SPS and SF slope for the first time, with a goal of connecting HI statistical properties with the underlying turbulent drivers. With future high-resolution HI observations these methods can be applied to many other galaxies. Having a high spatial resolution so we can reach down to the scales of stellar feedback sources (e.g. individual HII regions and SNe), as well as a high spatial dynamic range, is needed for mapping out spatial variations of interstellar turbulence. With the upcoming new radio telescopes like ASKAP, MeerKAT and the SKA such data sets will become increasingly more available.

\subsection{The SMC}

In the case of the SMC, we find highly uniform turbulent properties of HI. We have employed the rolling SPS, the rolling SF, and also investigated individual velocity channels where based on \cite{Offner15} turbulent properties should be more pronounced. However, we do not find evidence for local enhancements of turbulence due to stellar feedback.
A similar conclusion was reached in \cite{Nestingen-Palm17} who investigated turbulent properties within the central, star-forming SMC vs outer SMC regions and did not find any difference. Nestingen-Palm et al. (2017) showed that this lack of difference of turbulent properties can not be explained by the high optical depth HI, and that it is most likely due to a significant turbulent driving on large-scales. Numerical simulations by \cite{Yoo14} showed that the turbulent energy spectrum is very sensitive to large-scale driving. 
Unless the energy injection rates on small scales are much larger than the energy injection rate on large scales, the large-scale driving will always dominate.

Alternatively, we may be probing incorrect modes of turbulence. For example, several studies 
have shown that mechanical feedback from stellar winds is reduced in low metallicity environments \citep{Ramachandran19}. Therefore, supernova-driven turbulence could be more pronounced in the SMC than the LMC or the MW. \cite{Orkisz17} showed that in the case of Orion star forming region most of the cloud motions are  solenoidal, while around main star forming regions motions are driven by compressive turbulent forcing. As proposed by 
\cite{Federrath12}, the supernova-driven turbulence is likely to be more effective in producing compressive motions. Therefore undertaking statistical analysis that can separate solenoidal from compressive motions is an important future task.

Finally, the SMC is known to have a large line-of-sight depth, e.g. \cite{Mackey18}. It is therefore possible that effects of stellar feedback, which are localized in space and velocity, may simply get quickly washed out when integrating over a very long line-of-sight. To roughly test this hypothesis we have used our fBm simulations to create a random image with the SPS slope of $-3.4$. We then inserted inside this image a smaller sub-region having simulated a much steeper slope of $-4.0$ to effectively mimic the localized effect of stellar feedback. The sub-image had smoothed edges in order to blend well and not cause significant Gibbs ringing. We have then generated several additional images with only a slope of $-3.4$ and added all images together to effectively simulate a longer line-of-sight of the SMC. This experiment showed that adding only 2-3 images was able to wash out the spatial evidence of stellar feedback. While this is not a conclusive experiment and does not simulate velocity fluctuations produced by stellar feedback, it demonstrates that a significant line-of-sight depth could mask out localized regions with a steeper SPS slope. 

\subsection{The LMC}

In contrast to the SMC, the LMC HI shows a complex diversity in terms of its turbulent properties. We find that across most of the LMC, the small-scale slope is steeper than the large-scale slope. The break in the SPS of the LMC HI data from \cite{Kim98} has been first 
noticed by Elmegreen et al. (2001)
who found a slight steepening in the SPS on scales smaller than $\sim100$ pc and suggested that this could mark the transition from HI emission that is coming from a relatively thinner line-of-sight to that corresponding to the thicker line-of-sight. This line-of-sight depth change is caused by the change in the depth of the HI layer due to the disk structure of the LMC. 
In Elmegreen et al. (2001) the SPS slope on small scales was steep, about $-3.7$, but flattened on scales $>100-200$ pc to about $-2.7$.   Using 100 $\mu$m Spitzer observations of the LMC, Block et al. (2010) noticed a change in SPS slope from $-3.1$ on scales $<100-200$ pc while it became more shallow ($-2.1$) on larger scales. 
Bournaud et al. (2010) simulated an LMC-sized galaxy and reproduced the double power-law SPS of the HI column density. They found a power-law on large scales with a slope of $-1.9$, while on small scales the slope was $-3.1$. On scales smaller than the disk scaleheight, the gas density has the density power spectrum expected for a 3D turbulence. The motions corresponding to the large scales are quasi-2D and highly anisotropic. The SPS slopes we find are in agreement with Bournaud et al. (2010) simulations, as well as previous HI and infrared studies. 

\subsubsection{Influence of stellar feedback}

However, relative to previous studies we also find significant localized changes in the SPS slope. On small spatial scales, we find several areas of localized steepening of the SPS slope around major HII regions. The most prominent is the 30 Doradus region where the small-scale SPS slope changes from $\sim-3$ to $\sim-4$. This is in agreement with predictions from numerical simulations, e.g. 
\cite{Walker14} and Grisdale et al. (2017), where stellar feedback erodes and destroys small clouds, and effectively shifts power from small to large scales. This is the first direct observational evidence of localized modifications of the SPS caused by stellar feedback.
In the Grisdale et al. (2017) simulations stellar feedback affects the nature of turbulence by increasing the fraction of energy in compressive turbulent modes (via supernovae and HII region expansion and heating).
We note that the HI column density image of the LMC used in our study was not corrected for the high optical depth as HI absorption spectra in the direction of the LMC are currently not available (this will be possible in the near future especially with the upcoming ASKAP observations).

\subsubsection{Flaring of the LMC HI disk}

We also find localized large-scale steepening of the SPS slope in the outskirts of the LMC. We suggest that this is likely due to flaring of the HI disk and the increase of the HI thickness that causes transitions from 2D to 3D turbulence. Alternatively, the SPS steepening could be caused by the ram-pressure stripping on the South-East side, and/or magnetorotational instability which has been proposed to be important for driving turbulence in outer regions of galaxies \citep{Piontek05}.
While HI observational studies of the LMC have so far failed to detect flaring of the HI disk, numerical simulations have suggested that both stellar and gas disks should be significantly warped and flared due to gravitational interactions with the SMC.
\cite{Besla12}  performed numerical simulations of the SMC-LMC interactions with a goal of reproducing large-scale HI and stellar structure of the Magellanic System. In their models they found that the outer LMC stellar and gas disks are significantly warped and distorted. Their figures show significant warp of the gas disk on one side, due to interactions with the SMC, while on both sides the disk appears flared at about 5-6 kpc away from the center (roughly at about 6 degrees from the center).

The stellar LMC disk has been observed to be both warped and flared 
\citep{Alves00,vanderMarel01,Olsen02,Nikolaev04}.
By observing red clump stars in 50 randomly selected fields in the LMC, Olsen \& Salyk (2002) found that fields in the South-Western edge fall systematically low when compared with the fitted plane and concluded that the most likely explanation is the existence of a stellar warp. They suggested that the warp begins at a radius $\sim3$ degrees from the LMC center, is about 2 kpc wide, and lies out of the plane by $\sim2.5$ kpc. More recently using the Dark Energy Camera, 
\cite{Mackey18} discovered that the outer LMC stellar disk is significantly truncated both on the West and South (the side closest to the SMC) side, while being extended in the North. The North-South orientation of the outer disk is different from that of the inner disk, in agreement with the expectations if the disk is warped.

In the Milky Way HI disk, the locations of the HI warp and flare roughly coincide \citep{Levine06}.
The steepening of the SPS slope we find in the South-West
appears roughly in agreement (in terms of both position and extent) with the Olsen \& Salyk (2002) warp. Therefore, we conclude that the SPS slope change is most likely caused by the
transition from 2D to 3D turbulence due the increase of the HI disk thickness starting at 
the position of the stellar warp.  Although not well constrained by our fitting, the break point map suggests HI thickness $>150$ pc in this region.

Based on recent proper motion measurements, it appears that the complex morphology of the LMC disk is heavily influenced by interactions with the SMC. \cite{Choi18} used $\sim2.2$ million red clump stars from the Survey of the MAgellanic Stellar History (SMASH) to examine LMC stellar disk structure. 
They reproduced the Olsen \& Salyk (2002) inner warp in the South-West. Thanks to their large spatial coverage they also discovered a new prominent warp in the South-West that is located about 7 degrees from the center and departs from the fitted disk plane by $\sim4$ kpc. As Besla et al. (2012) predicted in one of their simulations (Model 2) outer warps both in stellar and gaseous disks, Choi et al. concluded that their warp is in agreement with predictions from simulations for the scenario of a recent direct collision between the SMC and LMC.
The Choi et al. warp is beyond the extent of the current HI data and remains to be examined in the future.

The HI disk of the LMC has been known to have an elongated region in the South-East that is aligned with the border of the optical disk. The region shows a steep increase in the HI column density distribution  \citep{Putman03,Staveley-Smith03}. Since the LMC proper motion vector is directed to the east, it has been suggested that this high density region was caused by ram pressure acting on the leading edge of the LMC disk and is therefore due to 
interactions with the hot MW halo. 
\cite{Mastropietro09} simulated the interaction between the MW halo and the LMC gaseous disk and showed that gaseous disk becomes strongly asymmetric with a compression at the front edge, in agreement with what is seen in HI. In their simulations, the high density region is well defined and localized within 1.5 kpc from the disk border. In addition, they find that the average thickness and velocity dispersion along the line of sight are larger in this feature than in the rest of the disk.
\cite{Salem15} simulated the implications of ram pressure on the LMC HI and showed that simulations can match the observed leading edge if the MW halo density is $\sim10^{-4}$ cm$^{-3}$. 
The elongated area of steep large-scale SPS slope we find in the South-East coincides with the leading HI edge where significant ram pressure effects are expected. With the current data we can not distinguish whether this steep slope area is tracing the flaring of the HI disk or a shock-like turbulence induced by ram pressure. Higher resolution, and larger area images, of the LMC in the future would allow us to map the SPS slope further beyond the stellar disk and properly fit the SPS slope break associated with the HI flare, therefore distinguishing between the two scenarios. 

\subsubsection{The LMC disk thickness}
 
Padoan et al. (2001) used the spectral correlation function to map out the spatial distribution of the HI scale height in the LMC. They found that the scale height varies from 130 pc to 280 pc, with the largest scale height being found around 30 Doradus. Their average HI scale height was in excellent agreement with Kim et al. (1998) who used the HI velocity dispersion to estimated the scale height of $\sim180$ pc. Interestingly, Padoan et al. (2001) also mentioned that sometimes on spatial scales of few hundred pc they found another possible change in the spectral correlation function and suggested that this may be caused by the effects of differential rotation. 
Throughout the LMC we find the disk thickness ranging from $\sim50$ to $\sim400$ pc. The 30 Doradus has the lowest disk thickness, while the elongated regions in the LMC outskirts have a disk thickness of $>150$ pc.


\section{Conclusions}

We developed methods for mapping spatial variations of the SPS and SF slope for the first time, with a goal of connecting HI statistical properties with the underlying turbulent drivers. 
The high spatial resolution provided by the ATCA and the ASKAP enabled us to reach down to the spatial scales where stellar feedback is predicted to modify significantly the turbulent HI spectrum. 
\begin{itemize}
\item With the rolling SPS and SF methods we find that HI in the SMC has highly uniform turbulent properties. This could be due to stellar feedback effects being  mitigated by low-metallicity, either by 
requiring even higher resolution observations as clouds get more compact or by our statistical methods not being able to isolate the exact turbulent modes at play. Alternatively, a large line-of-sight depth of the SMC could be washing out localized (both in space and velocity) effects of stellar feedback.
\item The rolling SPS method revealed significant spatial variations of the SPS slope across the LMC. Due to the HI disk structure, throughout most of the LMC the SPS requires a double power law fit. In agreement with previous studies, we find a more shallow slope on large spatial scales. On spatial scales smaller than 200-300 pc, we have discovered several regions with localized steepening of the SPS slope. The most prominent such region is the 30 Doradus. Such steepening of the SPS slope is predicted by numerical simulations as a result of stellar feedback eroding and destroying small clouds, effectively modifying the turbulent spectrum. Our results represent the first direct evidence for stellar feedback influencing the turbulent spectrum around large HII regions.
\item In the outskirts of the LMC, the rolling SPS method revealed two elongated, localized regions of a very steep SPS slope. The South-West region is at the position where stellar disk of the LMC has a warp. As numerical simulations suggest that both stellar and HI disks should be warped and flaring, the steep SPS slope is likely tracing the transition from 2D to 3D turbulence due to the increase of the HI disk scaleheight.
The South-East region coincides with the Leading edge of the LMC, where theoretical and numerical studies have suggested significant ram pressure effects as the LMC disk interacts with the MW halo.
\end{itemize}

S.S. acknowledges the support provided by the NSF Early Career Development (CAREER) Award AST-1056780, the Vilas funding provided by the University of Wisconsin, and the John Simon Guggenheim fellowship. This research was partially supported by the Munich Institute for Astro- and Particle Physics (MIAPP) of the DFG cluster of excellence ``Origin and Structure of the Universe''. N.Mc.-G. acknowledges funding from the Australian Research Council via grant FT150100024. 
We gratefully acknowledge contributions by W. Raja \& K. Bannister to ASKAP commissioning.
N.Mc.-G. also acknowledges support from the Diermeier visiting program  at the University of Wisconsin. The authors thank the anonymous referee for constructive comments and suggestions.
The authors are grateful to Daniel Eck, Diego Gonzalez Casanova, Eric Koch, Kearn Grisdale, Chad Bustard, Jay Gallagher, Bruce Elmegreen and Elena D'Onghia, for stimulating discussions.
The Australian SKA Pathfinder is part of the Australia Telescope National Facility which is managed by CSIRO. Operation of ASKAP is funded by the Australian Government with support from the National Collaborative Research Infrastructure Strategy. ASKAP uses the resources of the Pawsey Supercomputing Centre. Establishment of ASKAP, the Murchison Radio-astronomy Observatory and the Pawsey Super-computing Centre are initiatives of the Australian Government, with support from the Government of
Western Australia and the Science and Industry Endowment Fund. 
We acknowledge the Wajarri Yamatji
people as the traditional owners of the Observatory site. 
The The Australia Telescope Compact Array and the Parkes Radio Telescope are part of the Australia Telescope, which is funded by the Commonwealth of Australia for operation
as a National Facility managed by CSIRO. 
The Magellanic Clouds Emission Line Survey
(MCELS) data are provided by R. C. Smith, P. F. Winkler, and S. D. Points.  
The MCELS project has
been supported in part by NSF grants AST-9540747 and AST-0307613, and through 
the generous support of the Dean B. McLaughlin Fund at the University of Michigan, 
a bequest from the family of Dr. Dean
B. McLaughlin in memory of his lasting impact on Astronomy. 
This research made use of Astropy, a
community-developed core Python package for Astronomy 37. Parts of this research were conducted by the Australian Research Council Centre of Excellence for All Sky Astrophysics in 3 Dimensions (ASTRO 3D), through project number CE170100013. 

Astropy \citep{Astropy13}, MIRIAD \citep{Sault95}, KARMA \citep{Gooch96}, SciPy \citep{Oliphant07}, NumPy \citep{Walt11}, Matplotlib \citep{Hunter07}

\bibliographystyle{aasjournal}
\bibliography{rnaas}

\begin{thebibliography}{}
\expandafter\ifx\csname natexlab\endcsname\relax\def\natexlab#1{#1}\fi
\providecommand{\url}[1]{\href{#1}{#1}}
\providecommand{\dodoi}[1]{doi:~\href{http://doi.org/#1}{\nolinkurl{#1}}}
\providecommand{\doeprint}[1]{\href{http://ascl.net/#1}{\nolinkurl{http://ascl.net/#1}}}
\providecommand{\doarXiv}[1]{\href{https://arxiv.org/abs/#1}{\nolinkurl{https://arxiv.org/abs/#1}}}

\bibitem[{{Alves} \& {Nelson}(2000)}]{Alves00}
{Alves}, D.~R., \& {Nelson}, C.~A. 2000, \apj, 542, 789, \dodoi{10.1086/317023}

\bibitem[{{Astropy Collaboration} {et~al.}(2013){Astropy Collaboration},
  {Robitaille}, {Tollerud}, {Greenfield}, {Droettboom}, {Bray}, {Aldcroft},
  {Davis}, {Ginsburg}, {Price-Whelan}, {Kerzendorf}, {Conley}, {Crighton},
  {Barbary}, {Muna}, {Ferguson}, {Grollier}, {Parikh}, {Nair}, {Unther},
  {Deil}, {Woillez}, {Conseil}, {Kramer}, {Turner}, {Singer}, {Fox}, {Weaver},
  {Zabalza}, {Edwards}, {Azalee Bostroem}, {Burke}, {Casey}, {Crawford},
  {Dencheva}, {Ely}, {Jenness}, {Labrie}, {Lim}, {Pierfederici}, {Pontzen},
  {Ptak}, {Refsdal}, {Servillat}, \& {Streicher}}]{Astropy13}
{Astropy Collaboration}, {Robitaille}, T.~P., {Tollerud}, E.~J., {et~al.} 2013,
  \aap, 558, A33, \dodoi{10.1051/0004-6361/201322068}

\bibitem[{{Besla} {et~al.}(2012){Besla}, {Kallivayalil}, {Hernquist}, {van der
  Marel}, {Cox}, \& {Kere{\v s}}}]{Besla12}
{Besla}, G., {Kallivayalil}, N., {Hernquist}, L., {et~al.} 2012, \mnras, 421,
  2109, \dodoi{10.1111/j.1365-2966.2012.20466.x}

\bibitem[{{Block} {et~al.}(2010){Block}, {Puerari}, {Elmegreen}, \&
  {Bournaud}}]{Block10}
{Block}, D.~L., {Puerari}, I., {Elmegreen}, B.~G., \& {Bournaud}, F. 2010,
  \apjl, 718, L1, \dodoi{10.1088/2041-8205/718/1/L1}

\bibitem[{{Bolatto} {et~al.}(2011){Bolatto}, {Leroy}, {Jameson}, {Ostriker},
  {Gordon}, {Lawton}, {Stanimirovi{\'c}}, {Israel}, {Madden}, {Hony},
  {Sandstrom}, {Bot}, {Rubio}, {Winkler}, {Roman-Duval}, {van Loon},
  {Oliveira}, \& {Indebetouw}}]{Bolatto11}
{Bolatto}, A.~D., {Leroy}, A.~K., {Jameson}, K., {et~al.} 2011, \apj, 741, 12,
  \dodoi{10.1088/0004-637X/741/1/12}

\bibitem[{{Bournaud} {et~al.}(2010){Bournaud}, {Elmegreen}, {Teyssier},
  {Block}, \& {Puerari}}]{Bournaud10}
{Bournaud}, F., {Elmegreen}, B.~G., {Teyssier}, R., {Block}, D.~L., \&
  {Puerari}, I. 2010, \mnras, 409, 1088,
  \dodoi{10.1111/j.1365-2966.2010.17370.x}

\bibitem[{{Bracewell}(2006)}]{Bracewell06}
{Bracewell}, R.~N. 2006, Fourier analysis and imaging (Sprinegr)

\bibitem[{{Braun}(2012)}]{Braun12}
{Braun}, R. 2012, \apj, 749, 87, \dodoi{10.1088/0004-637X/749/1/87}

\bibitem[{{Burkhart} {et~al.}(2010){Burkhart}, {Stanimirovi{\'c}}, {Lazarian},
  \& {Kowal}}]{Burkhart10}
{Burkhart}, B., {Stanimirovi{\'c}}, S., {Lazarian}, A., \& {Kowal}, G. 2010,
  \apj, 708, 1204, \dodoi{10.1088/0004-637X/708/2/1204}

\bibitem[{{Calzetti} {et~al.}(2007){Calzetti}, {Kennicutt}, {Engelbracht},
  {Leitherer}, {Draine}, {Kewley}, {Moustakas}, {Sosey}, {Dale}, {Gordon},
  {Helou}, {Hollenbach}, {Armus}, {Bendo}, {Bot}, {Buckalew}, {Jarrett}, {Li},
  {Meyer}, {Murphy}, {Prescott}, {Regan}, {Rieke}, {Roussel}, {Sheth}, {Smith},
  {Thornley}, \& {Walter}}]{Calzetti07}
{Calzetti}, D., {Kennicutt}, R.~C., {Engelbracht}, C.~W., {et~al.} 2007, \apj,
  666, 870, \dodoi{10.1086/520082}

\bibitem[{{Choi} {et~al.}(2018){Choi}, {Nidever}, {Olsen}, {Blum}, {Besla},
  {Zaritsky}, {van der Marel}, {Bell}, {Gallart}, {Cioni}, {Johnson}, {Vivas},
  {Saha}, {de Boer}, {No{\"e}l}, {Monachesi}, {Massana}, {Conn},
  {Martinez-Delgado}, {Mu{\~n}oz}, \& {Stringfellow}}]{Choi18}
{Choi}, Y., {Nidever}, D.~L., {Olsen}, K., {et~al.} 2018, \apj, 866, 90,
  \dodoi{10.3847/1538-4357/aae083}

\bibitem[{{Combes} {et~al.}(2012){Combes}, {Boquien}, {Kramer}, {Xilouris},
  {Bertoldi}, {Braine}, {Buchbender}, {Calzetti}, {Gratier}, {Israel},
  {Koribalski}, {Lord}, {Quintana-Lacaci}, {Rela{\~n}o}, {R{\"o}llig},
  {Stacey}, {Tabatabaei}, {Tilanus}, {van der Tak}, {van der Werf}, \&
  {Verley}}]{Combes12}
{Combes}, F., {Boquien}, M., {Kramer}, C., {et~al.} 2012, \aap, 539, A67,
  \dodoi{10.1051/0004-6361/201118282}

\bibitem[{{Cornwell} {et~al.}(2011){Cornwell}, {Humphreys}, {Lenc}, {Voronkov},
  \& {Whiting}}]{Cornwell11}
{Cornwell}, T.~J., {Humphreys}, B., {Lenc}, E., {Voronkov}, M., \& {Whiting},
  M.~T. 2011, ASKAP Science Processing: Askap-sw-0020 Technical report 028
  (ASKAP Science Case Memo Series, CSIRO, 2011), CSIRO

\bibitem[{Crovisier \& Dickey(1983)}]{Crovisier83}
Crovisier, J., \& Dickey, J.~M. 1983, A$\&$A, 122, 282

\bibitem[{{de Avillez} \& {Breitschwerdt}(2005)}]{deAvillez05}
{de Avillez}, M.~A., \& {Breitschwerdt}, D. 2005, \aap, 436, 585,
  \dodoi{10.1051/0004-6361:20042146}

\bibitem[{Dickey {et~al.}(2000)Dickey, Mebold, Stanimirovic, \&
  Staveley-Smith}]{Dickey00}
Dickey, J.~M., Mebold, U., Stanimirovic, S., \& Staveley-Smith, L. 2000, \apj,
  536, 756

\bibitem[{{Dufour}(1984)}]{Dufour84}
{Dufour}, R.~J. 1984, in IAU Symposium, Vol. 108, Structure and Evolution of
  the Magellanic Clouds, ed. S.~{van den Bergh} \& K.~S.~D. {de Boer}, 353--360

\bibitem[{{Dutta} {et~al.}(2009{\natexlab{a}}){Dutta}, {Begum}, {Bharadwaj}, \&
  {Chengalur}}]{Dutta09a}
{Dutta}, P., {Begum}, A., {Bharadwaj}, S., \& {Chengalur}, J.~N.
  2009{\natexlab{a}}, \mnras, 397, L60,
  \dodoi{10.1111/j.1745-3933.2009.00684.x}

\bibitem[{{Dutta} {et~al.}(2009{\natexlab{b}}){Dutta}, {Begum}, {Bharadwaj}, \&
  {Chengalur}}]{Dutta09b}
---. 2009{\natexlab{b}}, \mnras, 398, 887,
  \dodoi{10.1111/j.1365-2966.2009.15105.x}

\bibitem[{Elmegreen(2000)}]{Elmegreen00}
Elmegreen, B.~G. 2000, 527, 266

\bibitem[{{Elmegreen} {et~al.}(2003){Elmegreen}, {Elmegreen}, \&
  {Leitner}}]{Elmegreen03}
{Elmegreen}, B.~G., {Elmegreen}, D.~M., \& {Leitner}, S.~N. 2003, \apj, 590,
  271, \dodoi{10.1086/374860}

\bibitem[{Elmegreen {et~al.}(2001)Elmegreen, Kim, \&
  Staveley-Smith}]{Elmegreen00a}
Elmegreen, B.~G., Kim, S., \& Staveley-Smith, L. 2001, 548, 749

\bibitem[{{Elmegreen} \& {Scalo}(2004)}]{Elmegreen04}
{Elmegreen}, B.~G., \& {Scalo}, J. 2004, \araa, 42, 211

\bibitem[{{Federrath} \& {Klessen}(2012)}]{Federrath12}
{Federrath}, C., \& {Klessen}, R.~S. 2012, \apj, 761, 156,
  \dodoi{10.1088/0004-637X/761/2/156}

\bibitem[{{Gaustad} {et~al.}(2001){Gaustad}, {McCullough}, {Rosing}, \& {Van
  Buren}}]{Gaustad01}
{Gaustad}, J.~E., {McCullough}, P.~R., {Rosing}, W., \& {Van Buren}, D. 2001,
  \pasp, 113, 1326, \dodoi{10.1086/323969}

\bibitem[{{Gooch}(1996)}]{Gooch96}
{Gooch}, R. 1996, in Astronomical Society of the Pacific Conference Series,
  Vol. 101, Astronomical Data Analysis Software and Systems V, ed. G.~H.
  {Jacoby} \& J.~{Barnes}, 80

\bibitem[{{Gordon} {et~al.}(2011){Gordon}, {Meixner}, {Meade}, {Whitney},
  {Engelbracht}, {Bot}, {Boyer}, {Lawton}, {Sewi{\l}o}, {Babler}, {Bernard},
  {Bracker}, {Block}, {Blum}, {Bolatto}, {Bonanos}, {Harris}, {Hora},
  {Indebetouw}, {Misselt}, {Reach}, {Shiao}, {Tielens}, {Carlson},
  {Churchwell}, {Clayton}, {Chen}, {Cohen}, {Fukui}, {Gorjian}, {Hony},
  {Israel}, {Kawamura}, {Kemper}, {Leroy}, {Li}, {Madden}, {Marble},
  {McDonald}, {Mizuno}, {Mizuno}, {Muller}, {Oliveira}, {Olsen}, {Onishi},
  {Paladini}, {Paradis}, {Points}, {Robitaille}, {Rubin}, {Sandstrom}, {Sato},
  {Shibai}, {Simon}, {Smith}, {Srinivasan}, {Vijh}, {Van Dyk}, {van Loon}, \&
  {Zaritsky}}]{Gordon11}
{Gordon}, K.~D., {Meixner}, M., {Meade}, M.~R., {et~al.} 2011, \aj, 142, 102,
  \dodoi{10.1088/0004-6256/142/4/102}

\bibitem[{Green(1993)}]{Green93}
Green, D.~A. 1993, 262, 327

\bibitem[{{Grisdale} {et~al.}(2018){Grisdale}, {Agertz}, {Renaud}, \&
  {Romeo}}]{Grisdale18}
{Grisdale}, K., {Agertz}, O., {Renaud}, F., \& {Romeo}, A.~B. 2018, \mnras,
  479, 3167, \dodoi{10.1093/mnras/sty1595}

\bibitem[{{Grisdale} {et~al.}(2019){Grisdale}, {Agertz}, {Renaud}, {Romeo},
  {Devriendt}, \& {Slyz}}]{Grisdale19}
{Grisdale}, K., {Agertz}, O., {Renaud}, F., {et~al.} 2019, \mnras, 486, 5482,
  \dodoi{10.1093/mnras/stz1201}

\bibitem[{{Grisdale} {et~al.}(2017){Grisdale}, {Agertz}, {Romeo}, {Renaud}, \&
  {Read}}]{Grisdale17}
{Grisdale}, K., {Agertz}, O., {Romeo}, A.~B., {Renaud}, F., \& {Read}, J.~I.
  2017, \mnras, 466, 1093, \dodoi{10.1093/mnras/stw3133}

\bibitem[{{HI4PI Collaboration} {et~al.}(2016){HI4PI Collaboration}, {Ben
  Bekhti}, {Fl{\"o}er}, {Keller}, {Kerp}, {Lenz}, {Winkel}, {Bailin},
  {Calabretta}, {Dedes}, {Ford}, {Gibson}, {Haud}, {Janowiecki}, {Kalberla},
  {Lockman}, {McClure-Griffiths}, {Murphy}, {Nakanishi}, {Pisano}, \&
  {Staveley-Smith}}]{HI4PI16}
{HI4PI Collaboration}, {Ben Bekhti}, N., {Fl{\"o}er}, L., {et~al.} 2016, \aap,
  594, A116, \dodoi{10.1051/0004-6361/201629178}

\bibitem[{Hou {et~al.}(1998)Hou, Wu, Chen, \& Zhou}]{Hou98}
Hou, T.~Y., Wu, X.-H., Chen, S., \& Zhou, Y. 1998, Phys. Rev. E, 58, 5841,
  \dodoi{10.1103/PhysRevE.58.5841}

\bibitem[{{Hunter}(2007)}]{Hunter07}
{Hunter}, J.~D. 2007, in Computing in Science \& Engineering, Vol.~9, 95

\bibitem[{{Jameson} {et~al.}(2016){Jameson}, {Bolatto}, {Leroy}, {Meixner},
  {Roman-Duval}, {Gordon}, {Hughes}, {Israel}, {Rubio}, {Indebetouw}, {Madden},
  {Bot}, {Hony}, {Cormier}, {Pellegrini}, {Galametz}, \&
  {Sonneborn}}]{Jameson16}
{Jameson}, K.~E., {Bolatto}, A.~D., {Leroy}, A.~K., {et~al.} 2016, \apj, 825,
  12, \dodoi{10.3847/0004-637X/825/1/12}

\bibitem[{{Jameson} {et~al.}(2019){Jameson}, {McClure-Griffiths}, {Liu},
  {Dickey}, {Staveley-Smith}, {Stanimirovi{\'c}}, {Dempsey}, {Dawson},
  {D{\'e}nes}, {Bolatto}, {Li}, \& {Wong}}]{Jameson19}
{Jameson}, K.~E., {McClure-Griffiths}, N.~M., {Liu}, B., {et~al.} 2019, \apjs,
  244, 7, \dodoi{10.3847/1538-4365/ab3576}

\bibitem[{{Joung} \& {Mac Low}(2006)}]{Joung06}
{Joung}, M.~K.~R., \& {Mac Low}, M.-M. 2006, \apj, 653, 1266,
  \dodoi{10.1086/508795}

\bibitem[{{Kim} {et~al.}(2001){Kim}, {Balsara}, \& {Mac Low}}]{Kim01}
{Kim}, J., {Balsara}, D., \& {Mac Low}, M.-M. 2001, Journal of Korean
  Astronomical Society, 34, 333, \dodoi{10.5303/JKAS.2001.34.4.333}

\bibitem[{{Kim} {et~al.}(1998){Kim}, {Staveley-Smith}, {Dopita}, {Freeman},
  {Sault}, {Kesteven}, \& {McConnell}}]{Kim98}
{Kim}, S., {Staveley-Smith}, L., {Dopita}, M.~A., {et~al.} 1998, \apj, 503,
  674, \dodoi{10.1086/306030}

\bibitem[{{Kim} {et~al.}(2003){Kim}, {Staveley-Smith}, {Dopita}, {Sault},
  {Freeman}, {Lee}, \& {Chu}}]{Kim03}
---. 2003, \apjs, 148, 473, \dodoi{10.1086/376980}

\bibitem[{{Krumholz} \& {Burkhart}(2016)}]{Krumholz16}
{Krumholz}, M.~R., \& {Burkhart}, B. 2016, \mnras, 458, 1671,
  \dodoi{10.1093/mnras/stw434}

\bibitem[{Lazarian \& Pogosyan(2000)}]{Lazarian99}
Lazarian, A., \& Pogosyan, D. 2000, 537, 720L

\bibitem[{{Lazarian} \& {Pogosyan}(2004)}]{Lazarian04}
{Lazarian}, A., \& {Pogosyan}, D. 2004, \apj, 616, 943, \dodoi{10.1086/422462}

\bibitem[{{Levine} {et~al.}(2006){Levine}, {Blitz}, \& {Heiles}}]{Levine06}
{Levine}, E.~S., {Blitz}, L., \& {Heiles}, C. 2006, \apj, 643, 881,
  \dodoi{10.1086/503091}

\bibitem[{{Mackey} {et~al.}(2018){Mackey}, {Koposov}, {Da Costa}, {Belokurov},
  {Erkal}, \& {Kuzma}}]{Mackey18}
{Mackey}, D., {Koposov}, S., {Da Costa}, G., {et~al.} 2018, \apjl, 858, L21,
  \dodoi{10.3847/2041-8213/aac175}

\bibitem[{{Maier} {et~al.}(2016){Maier}, {Chien}, \& {Hunter}}]{Maier16}
{Maier}, E., {Chien}, L.-H., \& {Hunter}, D.~A. 2016, \aj, 152, 134,
  \dodoi{10.3847/0004-6256/152/5/134}

\bibitem[{{Mastropietro} {et~al.}(2009){Mastropietro}, {Burkert}, \&
  {Moore}}]{Mastropietro09}
{Mastropietro}, C., {Burkert}, A., \& {Moore}, B. 2009, \mnras, 399, 2004,
  \dodoi{10.1111/j.1365-2966.2009.15406.x}

\bibitem[{{McClure-Griffiths} {et~al.}(2018){McClure-Griffiths}, {D{\'e}nes},
  {Dickey}, {Stanimirovi{\'c}}, {Staveley-Smith}, {Jameson}, {Di Teodoro},
  {Allison}, {Collier}, {Chippendale}, {Franzen}, {G{\"u}rkan}, {Heald},
  {Hotan}, {Kleiner}, {Lee-Waddell}, {McConnell}, {Popping}, {Rhee}, {Riseley},
  {Voronkov}, \& {Whiting}}]{McClure-Griffiths18}
{McClure-Griffiths}, N.~M., {D{\'e}nes}, H., {Dickey}, J.~M., {et~al.} 2018,
  Nature Astronomy, 2, 901, \dodoi{10.1038/s41550-018-0608-8}

\bibitem[{{Miville-Desch{\^e}nes} {et~al.}(2003){Miville-Desch{\^e}nes},
  {Levrier}, \& {Falgarone}}]{Miville-Deschenes03}
{Miville-Desch{\^e}nes}, M.-A., {Levrier}, F., \& {Falgarone}, E. 2003, \apj,
  593, 831, \dodoi{10.1086/376603}

\bibitem[{{Muller} {et~al.}(2004){Muller}, {Stanimirovi{\'c}}, {Rosolowsky}, \&
  {Staveley-Smith}}]{Muller04}
{Muller}, E., {Stanimirovi{\'c}}, S., {Rosolowsky}, E., \& {Staveley-Smith}, L.
  2004, \apj, 616, 845, \dodoi{10.1086/425154}

\bibitem[{{Nayak} {et~al.}(2019){Nayak}, {Meixner}, {Sewi{\l}o}, {Ochsendorf},
  {Bolatto}, {Indebetouw}, {Kawamura}, {Onishi}, \& {Fukui}}]{Nayak19}
{Nayak}, O., {Meixner}, M., {Sewi{\l}o}, M., {et~al.} 2019, \apj, 877, 135,
  \dodoi{10.3847/1538-4357/ab1b38}

\bibitem[{{Nestingen-Palm} {et~al.}(2017){Nestingen-Palm}, {Stanimirovi{\'c}},
  {Gonz{\'a}lez-Casanova}, {Babler}, {Jameson}, \&
  {Bolatto}}]{Nestingen-Palm17}
{Nestingen-Palm}, D., {Stanimirovi{\'c}}, S., {Gonz{\'a}lez-Casanova}, D.~F.,
  {et~al.} 2017, \apj, 845, 53, \dodoi{10.3847/1538-4357/aa7e78}

\bibitem[{{Nikolaev} {et~al.}(2004){Nikolaev}, {Drake}, {Keller}, {Cook},
  {Dalal}, {Griest}, {Welch}, \& {Kanbur}}]{Nikolaev04}
{Nikolaev}, S., {Drake}, A.~J., {Keller}, S.~C., {et~al.} 2004, \apj, 601, 260,
  \dodoi{10.1086/380439}

\bibitem[{{Offner} \& {Arce}(2015)}]{Offner15}
{Offner}, S.~S.~R., \& {Arce}, H.~G. 2015, \apj, 811, 146,
  \dodoi{10.1088/0004-637X/811/2/146}

\bibitem[{{Oliphant}(2007)}]{Oliphant07}
{Oliphant}, T.~E. 2007, in Computing in Science \& Engineering, Vol.~9, 20

\bibitem[{{Olsen} \& {Salyk}(2002)}]{Olsen02}
{Olsen}, K.~A.~G., \& {Salyk}, C. 2002, \aj, 124, 2045, \dodoi{10.1086/342739}

\bibitem[{{Orkisz} {et~al.}(2017){Orkisz}, {Pety}, {Gerin}, {Bron},
  {Guzm{\'a}n}, {Bardeau}, {Goicoechea}, {Gratier}, {Le Petit}, {Levrier},
  {Liszt}, {{\"O}berg}, {Peretto}, {Roueff}, {Sievers}, \&
  {Tremblin}}]{Orkisz17}
{Orkisz}, J.~H., {Pety}, J., {Gerin}, M., {et~al.} 2017, \aap, 599, A99,
  \dodoi{10.1051/0004-6361/201629220}

\bibitem[{{Padoan} {et~al.}(2001){Padoan}, {Kim}, {Goodman}, \&
  {Staveley-Smith}}]{Padoan01}
{Padoan}, P., {Kim}, S., {Goodman}, A., \& {Staveley-Smith}, L. 2001, \apjl,
  555, L33, \dodoi{10.1086/321735}

\bibitem[{{Pingel} {et~al.}(2018){Pingel}, {Lee}, {Burkhart}, \&
  {Stanimirovi{\'c}}}]{Pingel18}
{Pingel}, N.~M., {Lee}, M.-Y., {Burkhart}, B., \& {Stanimirovi{\'c}}, S. 2018,
  \apj, 856, 136, \dodoi{10.3847/1538-4357/aab34b}

\bibitem[{{Pingel} {et~al.}(2013){Pingel}, {Stanimirovi{\'c}}, {Peek}, {Lee},
  {Lazarian}, {Burkhart}, {Begum}, {Douglas}, {Heiles}, {Gibson}, {Grcevich},
  {Korpela}, {Lawrence}, {Murray}, {Putman}, \& {Saul}}]{Pingel13}
{Pingel}, N.~M., {Stanimirovi{\'c}}, S., {Peek}, J.~E.~G., {et~al.} 2013, \apj,
  779, 36, \dodoi{10.1088/0004-637X/779/1/36}

\bibitem[{{Piontek} \& {Ostriker}(2005)}]{Piontek05}
{Piontek}, R.~A., \& {Ostriker}, E.~C. 2005, \apj, 629, 849,
  \dodoi{10.1086/431549}

\bibitem[{{Putman} {et~al.}(2003){Putman}, {Staveley-Smith}, {Freeman},
  {Gibson}, \& {Barnes}}]{Putman03}
{Putman}, M.~E., {Staveley-Smith}, L., {Freeman}, K.~C., {Gibson}, B.~K., \&
  {Barnes}, D.~G. 2003, \apj, 586, 170

\bibitem[{{Ramachandran} {et~al.}(2019){Ramachandran}, {Hamann}, {Oskinova},
  {Gallagher}, {Hainich}, {Shenar}, {Sander}, {Todt}, \&
  {Fulmer}}]{Ramachandran19}
{Ramachandran}, V., {Hamann}, W.-R., {Oskinova}, L.~M., {et~al.} 2019, \aap,
  625, A104, \dodoi{10.1051/0004-6361/201935365}

\bibitem[{{Russell} \& {Dopita}(1992)}]{Russell92}
{Russell}, S.~C., \& {Dopita}, M.~A. 1992, \apj, 384, 508,
  \dodoi{10.1086/170893}

\bibitem[{{Salem} {et~al.}(2015){Salem}, {Besla}, {Bryan}, {Putman}, {van der
  Marel}, \& {Tonnesen}}]{Salem15}
{Salem}, M., {Besla}, G., {Bryan}, G., {et~al.} 2015, \apj, 815, 77,
  \dodoi{10.1088/0004-637X/815/1/77}

\bibitem[{{Sault} {et~al.}(1995){Sault}, {Teuben}, \& {Wright}}]{Sault95}
{Sault}, R.~J., {Teuben}, P.~J., \& {Wright}, M.~C.~H. 1995, in Astronomical
  Society of the Pacific Conference Series, Vol.~77, Astronomical Data Analysis
  Software and Systems IV, ed. R.~A. {Shaw}, H.~E. {Payne}, \& J.~J.~E.
  {Hayes}, 433

\bibitem[{{Simonetti} {et~al.}(1984){Simonetti}, {Cordes}, \&
  {Spangler}}]{Simonetti84}
{Simonetti}, J.~H., {Cordes}, J.~M., \& {Spangler}, S.~R. 1984, \apj, 284, 126,
  \dodoi{10.1086/162391}

\bibitem[{{Smith} \& {MCELS Team}(1999)}]{Smith99}
{Smith}, R.~C., \& {MCELS Team}. 1999, in IAU Symposium, Vol. 190, New Views of
  the Magellanic Clouds, ed. Y.-H. {Chu}, N.~{Suntzeff}, J.~{Hesser}, \&
  D.~{Bohlender}, 28

\bibitem[{{Snedecor} \& {Cochran}(1989)}]{Snedecor89}
{Snedecor}, G.~W., \& {Cochran}, W.~G. 1989, Statistical Methods (Iowa State
  University Press), 343

\bibitem[{{Stanimirovi{\'c}} \& {Lazarian}(2001)}]{Stanimirovic01}
{Stanimirovi{\'c}}, S., \& {Lazarian}, A. 2001, \apjl, 551, L53

\bibitem[{Stanimirovi\'{c} {et~al.}(1999)Stanimirovi\'{c}, Staveley-Smith,
  Dickey, Sault, \& Snowden}]{Stanimirovic99}
Stanimirovi\'{c}, S., Staveley-Smith, L., Dickey, J.~M., Sault, R.~J., \&
  Snowden, S.~L. 1999, 302, 417

\bibitem[{{Stanimirovi{\'c}} {et~al.}(2004){Stanimirovi{\'c}},
  {Staveley-Smith}, \& {Jones}}]{Stanimirovic04}
{Stanimirovi{\'c}}, S., {Staveley-Smith}, L., \& {Jones}, P.~A. 2004, \apj,
  604, 176, \dodoi{10.1086/381869}

\bibitem[{Stanimirovi\'{c} {et~al.}(2000)Stanimirovi\'{c}, Staveley-Smith,
  van~der Hulst, Bontekoe, Kester, \& Jones}]{Stanimirovic00}
Stanimirovi\'{c}, S., Staveley-Smith, L., van~der Hulst, J.~M., {et~al.} 2000,
  315, 791

\bibitem[{{Staveley-Smith} {et~al.}(2003){Staveley-Smith}, {Kim}, {Calabretta},
  {Haynes}, \& {Kesteven}}]{Staveley-Smith03}
{Staveley-Smith}, L., {Kim}, S., {Calabretta}, M.~R., {Haynes}, R.~F., \&
  {Kesteven}, M.~J. 2003, \mnras, 339, 87,
  \dodoi{10.1046/j.1365-8711.2003.06146.x}

\bibitem[{{Stilp} {et~al.}(2013){Stilp}, {Dalcanton}, {Skillman}, {Warren},
  {Ott}, \& {Koribalski}}]{Stilp13}
{Stilp}, A.~M., {Dalcanton}, J.~J., {Skillman}, E., {et~al.} 2013, \apj, 773,
  88, \dodoi{10.1088/0004-637X/773/2/88}

\bibitem[{{Tamburro} {et~al.}(2009){Tamburro}, {Rix}, {Leroy}, {Mac Low},
  {Walter}, {Kennicutt}, {Brinks}, \& {de Blok}}]{Tamburro09}
{Tamburro}, D., {Rix}, H.-W., {Leroy}, A.~K., {et~al.} 2009, \aj, 137, 4424,
  \dodoi{10.1088/0004-6256/137/5/4424}

\bibitem[{{van der Marel} \& {Cioni}(2001)}]{vanderMarel01}
{van der Marel}, R.~P., \& {Cioni}, M.-R.~L. 2001, \aj, 122, 1807,
  \dodoi{10.1086/323099}

\bibitem[{{van der Walt} {et~al.}(2011){van der Walt}, {Colbert}, \&
  {Varoquaux}}]{Walt11}
{van der Walt}, S., {Colbert}, S.~C., \& {Varoquaux}, G. 2011, in Computing in
  Science \& Engineering, Vol.~13, 30

\bibitem[{{Wada} {et~al.}(2002){Wada}, {Meurer}, \& {Norman}}]{Wada02}
{Wada}, K., {Meurer}, G., \& {Norman}, C.~A. 2002, \apj, 577, 197,
  \dodoi{10.1086/342151}

\bibitem[{{Walker} {et~al.}(2014){Walker}, {Gibson}, {Pilkington}, {Brook},
  {Dutta}, {Stanimirovi{\'c}}, {Stinson}, \& {Bailin}}]{Walker14}
{Walker}, A.~P., {Gibson}, B.~K., {Pilkington}, K., {et~al.} 2014, \mnras, 441,
  525, \dodoi{10.1093/mnras/stu419}

\bibitem[{{Warren} {et~al.}(2012){Warren}, {Skillman}, {Stilp}, {Dalcanton},
  {Ott}, {Walter}, {Petersen}, {Koribalski}, \& {West}}]{Warren12}
{Warren}, S.~R., {Skillman}, E.~D., {Stilp}, A.~M., {et~al.} 2012, \apj, 757,
  84, \dodoi{10.1088/0004-637X/757/1/84}

\bibitem[{{Yoo} \& {Cho}(2014)}]{Yoo14}
{Yoo}, H., \& {Cho}, J. 2014, \apj, 780, 99, \dodoi{10.1088/0004-637X/780/1/99}

\bibitem[{{Young} \& {Lo}(1996)}]{Young96}
{Young}, L.~M., \& {Lo}, K.~Y. 1996, \apj, 462, 203, \dodoi{10.1086/177141}

\bibitem[{{Zhang} {et~al.}(2012){Zhang}, {Hunter}, \& {Elmegreen}}]{Zhang12}
{Zhang}, H.-X., {Hunter}, D.~A., \& {Elmegreen}, B.~G. 2012, \apj, 754, 29,
  \dodoi{10.1088/0004-637X/754/1/29}

\end{thebibliography}

\appendix

\section{APPENDIX A: Spatial Power Spectrum Simulations}
\begin{figure} [h!]
  \centering
  \includegraphics[scale=0.7]{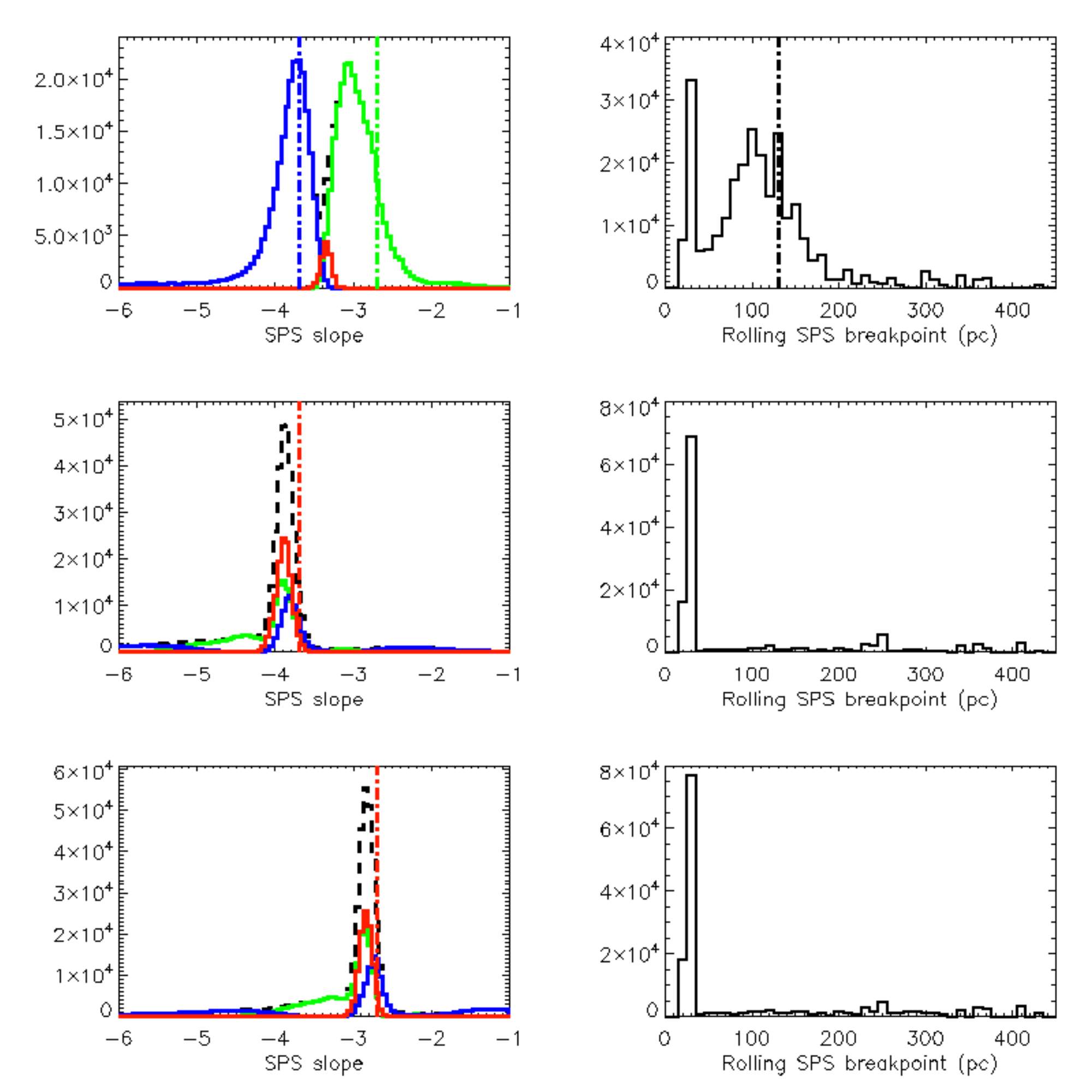}
  \caption{Results of the rolling SPS method when applied on fBm simulated images. Each simulated image was a size of 655$\times$655 pixels. A rolling SPS routine was performed with a kernel size of 128 pixels, shifted by 1 pixel. A single and double power-law fits were then calculated along with an F-Test of the significance between the 2 fits. The f-Test cutoff value was set at 0.05 significance. (Top) Simulated image with a double power-law, with a slope of $-3.7$ on small scales and a slope of $-2.7$ on large scales,  with the break point corresponding to 130 pc (or 9 pixels, we use a conversion of 1 pixel corresponding to 14.5 pc).
  The left column shows the results of the rolling SPS calculation:  estimated small-scale slope in blue, large-scale slopes in green, and in red the single slope values when the F-Test indicated a single slope solution is preferred.  Dash-dot lines show the input slope values.  The right column shows the estimated break point values when the F-Test preferred a 2-slope solution.  
  (Middle) Simulated image with a single power law of
  $-3.7$.
  (Bottom) Simulated image with a single power law of $-2.7$.
}
  \label{fig:Rolling_SPS_Sim_fTest}
\end{figure}

To effectively gauge the ability of the rolling SPS method to discern differing power indices, we used fBm simulated images.  We ran the rolling SPS routine on both single power law simulated images with $\gamma=-2.7$ and $-3.7$, but also a simulated image constructed of a broken power law with $\gamma=-3.7$ for small scales and $-2.7$ at the large scale end, where the break point in the power law corresponded to a scale of 9 pixels.  Simulated images were 655 by 655 pixels in size.  This broken power law image was constructed to simulate the HI SPS of the LMC as found by Elmegreen et al. (2001).  In Figure~\ref{fig:Rolling_SPS_Sim_fTest} we present the effectiveness of our rolling SPS method with the F-test to reproduce the SPS index value, as well as the break point location, when applied to 
simulated images.

  The top row of Figure~\ref{fig:Rolling_SPS_Sim_fTest} shows results for a simulated image with a double power-law, while the middle and bottom rows show results from single-slope simulated images, with $\gamma=-3.7$ and $-2.7$, respectively.
  The left column shows the results of the rolling SPS calculation:  estimated small-scale slope in blue, large-scale slopes in green, and in red the single slope values when the F-Test indicated a single slope solution is preferred.  Dash-dot lines show the input slope values.  The right column shows the estimated break point values when the F-Test preferred a 2-slope solution.

There are two main conclusions from these tests. First, the F-test correctly decides that the simulated double power-law image (top row) has two slopes, while two single power-law simulated images are found to have predominantly single slopes. For single power law simulated data (middle and bottom rows) we see that the preferred break point is typically limited to the shortest possible break point.  However, we see a large increase in the number of preferred single slope solutions (red).  The dashed black line in the second and third row shows the sum of the 3 histograms (blue, green and red)  
to illustrate that the overall preferred solution is near the input slope value.

\begin{figure}[h!]
  \centering
 \includegraphics[width=0.5\textwidth]{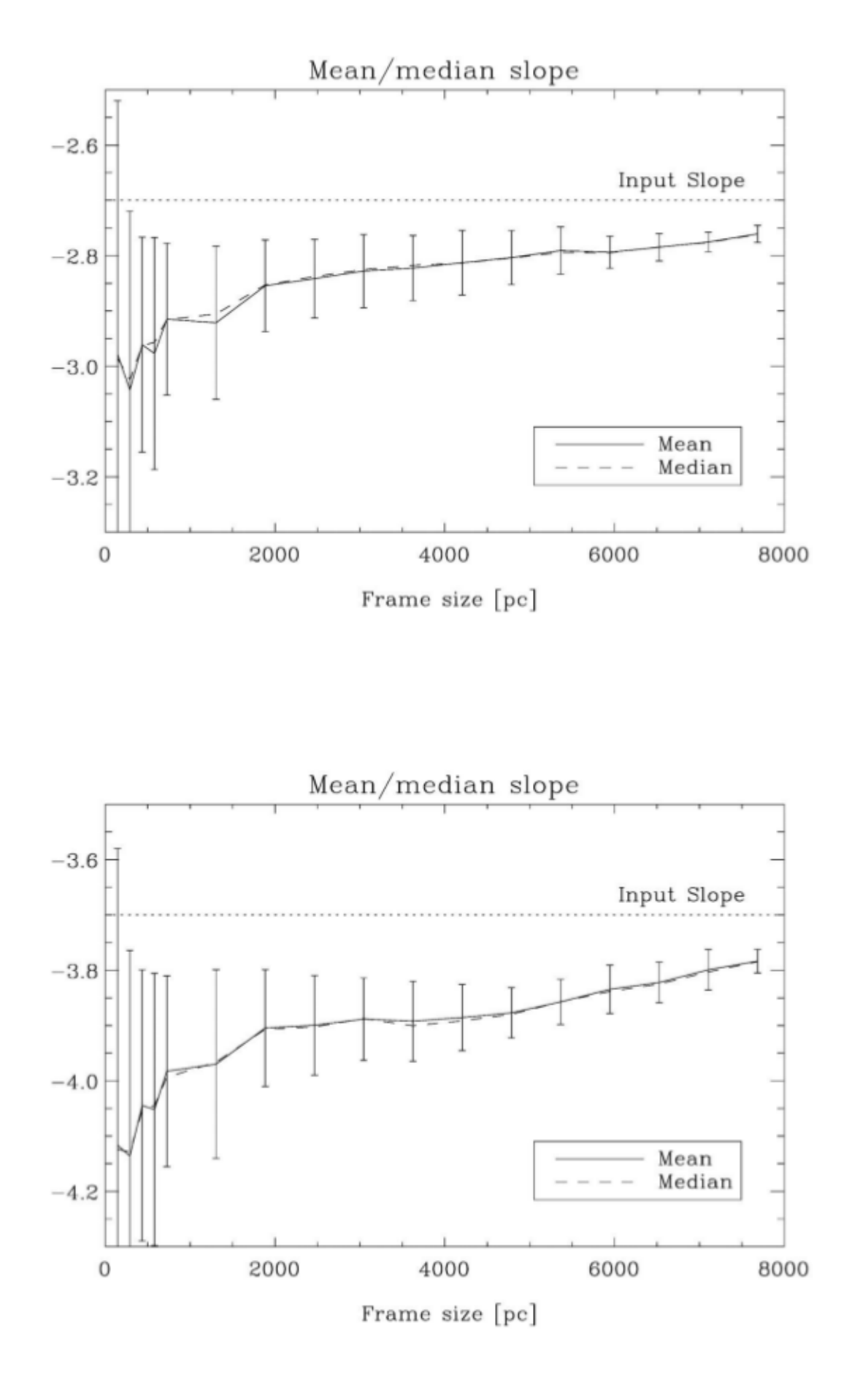}
   \label{fig:meanSlopevsImageSize}
   \caption{Example of how the kernel size affects calculations of the SPS slope. Two simulated fBm images are shown with the input SPS slope of $-2.7$ (top) and $-3.7$ (bottom). The x-axis shows the kernel size used to calculate the SPS slope. The kernel size varies from 500 pc to the full image size. By applying a given kernel size across the whole image we estimate the median/mean SPS slope. As the kernel size increases the recovered SPS slope from the rolling SPS method approaches the true input SPS slope.}
\end{figure}

Second, in all three cases the recovered slopes are slightly underestimated (steeper than the original, input slope). This is a consequence of the Gibbs effect on individual kernels which affects more larger scales, resulting  in a steeper slope.
Figure~\ref{fig:meanSlopevsImageSize} shows that as the kernel size increases and approaches the 655$\times$655 pixel image size, the standard deviation in these results decreases and the slope distribution peak shifts closer to the input values.
By doing these tests we conclude that our estimated slopes, for our kernel size of 1.5 kpc, are systematically underestimated by $\sim0.2$.

\begin{figure}[h!]
  \centering
  \includegraphics[width=0.95\textwidth]{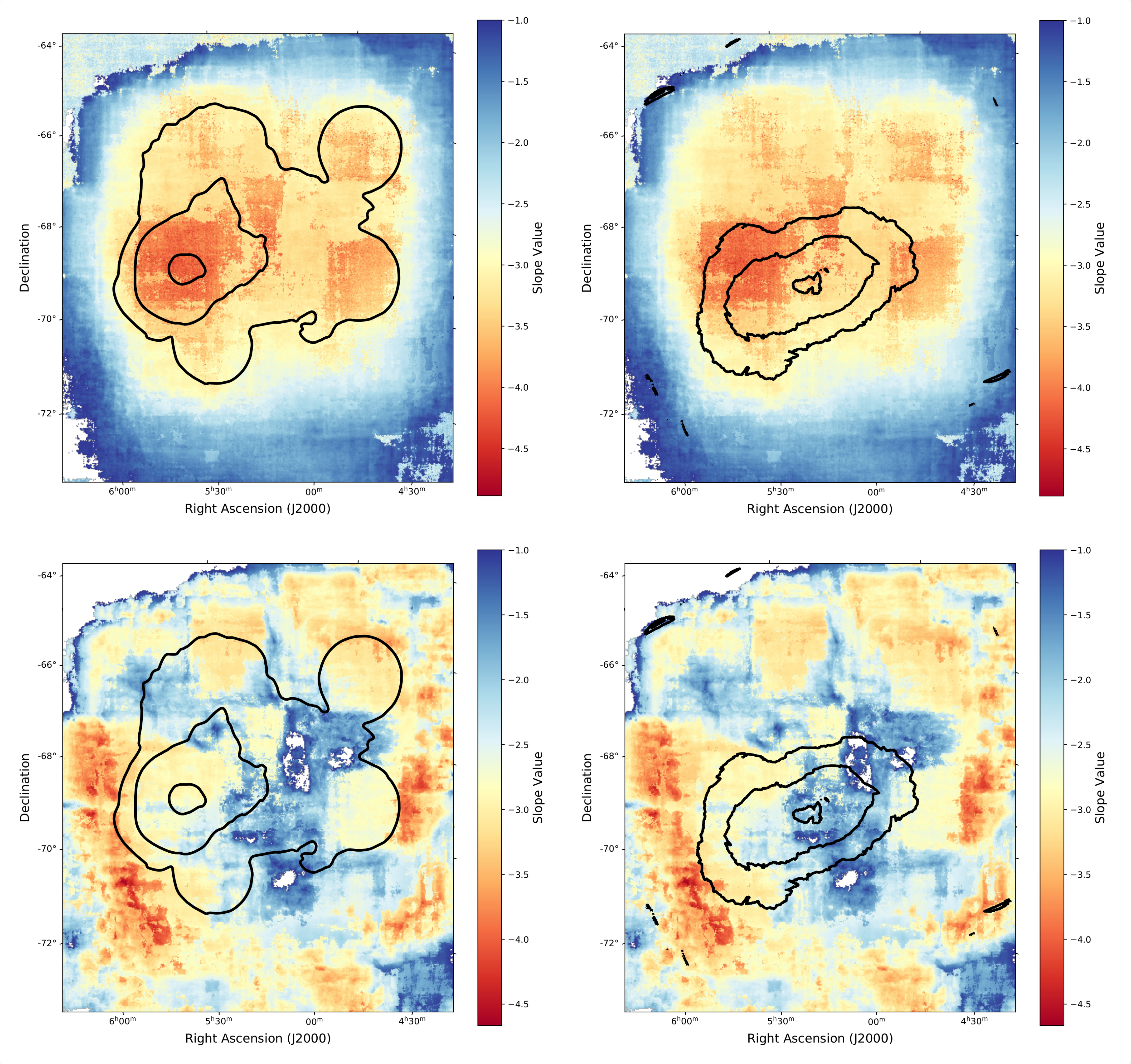}
  \caption{(Upper left) Map of the small scale SPS slope $\gamma$ across the integrated LMC image overlaid with the SFR surface density. (Upper right) Map of the small scale SPS slope $\gamma$ across the integrated LMC image overlaid with the stellar surface density. (Lower left) Map of the large scale SPS slope $\gamma$ across the integrated LMC image overlaid with the SFR surface density. (Lower left) Map of the large scale SPS slope $\gamma$ across the integrated LMC image overlaid with the stellar surface density. The SFR surface density contours are shown at 0.0022, 0.009, and 0.095 MJy/sr. The stellar surface density contours are at 0.25, 0.41, and 0.6 MJy/sr.}
\label{fig:gamma_large_small_SFR_stellar}
\end{figure}

\end{document}